\documentclass[aps,prd,twocolumn,eqsecnum,superscriptaddress,nofootinbib,letterpaper]{revtex4}

\usepackage{verbatim}
\usepackage[usenames,dvipsnames]{color}
\usepackage[pdftex]{graphicx}
\usepackage[english]{babel}
\usepackage{amsmath,amssymb,amsfonts}
\usepackage[colorlinks=true,citecolor=blue]{hyperref}

\usepackage[T1]{fontenc}
\usepackage{mathptmx}

\begin{document}
\newcommand{\fan}[1]{\textcolor{Blue}{#1}}
\newcommand{\david}[1]{\textcolor{PineGreen}{#1}}
\newcommand{\aaron}[1]{\textcolor{cyan}{#1}}
\newcommand{\R}[1]{\textcolor{WildStrawberry}{#1}}

\newcommand{\mycomments}[1]{}

\newcommand{\bsalign}
{
\begin{align}
}

\newcommand{\esalign}{\end{align}}

\newcommand{\E}{\mathrm{E}}
\newcommand{\Var}{\mathrm{Var}}
\newcommand{\bra}[1]{\langle #1|}
\newcommand{\ket}[1]{|#1\rangle}
\newcommand{\braket}[2]{\langle #1|#2 \rangle}
\newcommand{\mean}[2]{\langle #1 #2 \rangle}
\newcommand{\be}{\begin{equation}}
\newcommand{\ee}{\end{equation}}
\newcommand{\ba}{\begin{eqnarray}}
\newcommand{\ea}{\end{eqnarray}}
\newcommand{\SD}[1]{{\color{magenta}#1}}
\newcommand{\rem}[1]{{\sout{#1}}}
\newcommand{\alert}[1]{\textbf{\color{red} \uwave{#1}}}
\newcommand{\Y}[1]{\textcolor{yellow}{#1}}
\newcommand{\B}[1]{\textcolor{blue}{#1}}
\newcommand{\C}[1]{\textcolor{cyan}{#1}}
\newcommand{\db}{\color{darkblue}}
\newcommand{\intinfty}{\int_{-\infty}^{\infty}\!}
\newcommand{\Tr}{\mathop{\rm Tr}\nolimits}
\newcommand{\const}{\mathop{\rm const}\nolimits}
\makeatletter
\newcommand{\rmnum}[1]{\romannumeral #1}
\newcommand{\Rmnum}[1]{\expandafter\@slowromancap\romannumeral #1@}
\makeatother

\title{First-Order Perturbative Hamiltonian Equations of Motion for a Point Particle Orbiting a Schwarzschild Black Hole}
\author{Huan Yang}
\affiliation{Theoretical Astrophysics 350-17, California Institute of Technology, Pasadena, CA 91125, USA}
\author{Haixing Miao}
\affiliation{Theoretical Astrophysics 350-17, California Institute of Technology, Pasadena, CA 91125, USA}
\author{Yanbei Chen}
\affiliation{Theoretical Astrophysics 350-17, California Institute of Technology, Pasadena, CA 91125, USA}
\date{\today}


\begin{abstract}
We formulate a spherical harmonically decomposed 1+1 scheme to self-consistently evolve the trajectory of a point particle and its gravitational metric perturbation to a Schwarzschild background spacetime. 
Following the work of Moncrief, we write down an action for perturbations in space-time geometry, combine that with the action for a point-particle moving through this space-time, and then  obtain Hamiltonian equations of motion for metric perturbations, the particle's coordinates, as well as their canonical momenta. 
Hamiltonian equations for the metric-perturbation and their conjugate momenta, for even and odd parities, reduces to Zerilli-Moncrief and Regge-Wheeler master equations with source terms, which are gauge invariant, plus auxiliary equations that specify gauge. 
Hamiltonian equations for the particle, on the other hand,  now include effect of metric perturbations --- with these new terms derived from the same interaction Hamiltonian that had lead to those well-known source terms. 
In this way, space-time geometry and particle motion can be evolved in a self-consistent manner, in principle in any gauge. 
However, the point-particle nature of our source requires regularization, and we outline how the Detweiler-Whiting approach can be applied.
In this approach, a singular field can be obtained analytically using Hadamard decomposition of the Green's function and the regular field, which needs to be evolved numerically, is the result of subtracting the singular field from the total metric perturbation.  In principle, any gauge that has the singular-regular field decomposition is suitable for our self-consistent scheme. 
In reality, however, this freedom is only possible if our singular field has a high enough level of smoothness.  For a singular field with minimum quality, one can adopt the Lorenz gauge condition, which we have recast into our formalism: for each $l$ and $m$, we have 2 wave equations to evolve odd and even parity gauge invariant quantities and 8 first order differential equations to fix the Lorenz gauge and determine the metric components.
\end{abstract}

\maketitle

\section{introduction}

In this paper, we discuss the motion of a small compact object (idealizing a black hole or neutron star) moving around a much more massive, Schwarzschild black hole --- and the gravitational waves such a system would emit.  In gravitational-wave astrophysics, this process is often referred to as an Extreme Mass-Ratio Inspiral (EMRI).  This problem has attracted a lot of attention in recent years due to the possibility of directly detecting these waves using space-based ~\cite{abra,hughes,flanagan} and even ground-based laser interferometer gravitational-wave detectors~\cite{brown}. In EMRIs, the small object stays for a long time outside of the big black hole, emitting many cycles of gravitational waves --- even in the strong-field region very close to the big black hole.  This wave, if detected, will provide accurate information about the mass and the spin of big black hole, as well as parameters of the small object's orbit; one may even test whether the big black hole is indeed a Kerr background spacetime \cite{finn,jeandrew,collins,vigeland}. On the other hand, because it is the many cycles that would together lead to a detectable signal, it will be crucial (at least for the weaker sources) to get very accurate models for the waveforms (e.g., at the level of one or a few radians for the entire waveform, which may be up to $10^6$ cycles) in order to be able to extract them from data \cite{curt}. 

Because the orbiting object is much smaller in mass, one expects the application of black-hole perturbation theory \cite{regge,zerilli,teukolsky}, successively to higher orders in mass ratio, would be a viable program towards solving the EMRI problem, just like Post-Newtonian theory has worked for the inspiral of comparable-mass compact objects \cite{blanchet}--- although a direct application of post-Newtonian theory to EMRIs will not be very efficient because one expects the most interesting waves to be generated when the small object moves very close to the big black hole, with  where post-Newtonian theory breaks down very fast.  By contrast, full numerical simulation of the entire EMRI spacetime would be very expensive due to the large separation of scales and very long integration time that is required to providing meaningful information regarding the evolution of the orbit.  Nevertheless, the most extreme mass ratio achieved so far in numerical relativity simulations is $1:100$ \cite{lousto}.

When computing the leading-order waveform emitted by a small compact object moving in a black-hole background, one can idealize the small object as a test particle moving along a geodesic, and perturbations to the black-hole spacetime is sourced by a $\delta$-function stress-energy tensor along that geodesic --- with all other multipoles of the object ignored.   However, if we would like to further calculate the evolution of the object via coupling to the radiation field, we need to consider space-time geometry near the object, which formally diverges if we still use the point-particle model.  From this point of view, a regularization procedure is necessary.

Historically there are two approaches toward regularization. The first involves matching the external (point-particle-generated perturbated black-hole) spacetime to the internal (perturbed Schwarzschild) spacetime of the small object --- in a region where both are valid.  This was proposed and carried out by Mino, Sasaki and Tanaka~\cite{mino} as well as by Quinn and Wald \cite{quinn},  and later by Gralla, Pound, Poisson and others \cite{gralla, adam, poisson}. This approach, for the external spacetime of the object, has led to the separation of the total metric perturbation field into two pieces: $h=h_{\rm dir}+h_{\rm tail}$;  $h_{\rm dir}$ is the light-cone contribution to metric perturbation from the point particle's $\delta$-function stress energy tensor (the Hadamard direct part) and $h_{\rm tail}$ is the contribution inside the future light cone of the particle (the tail part). 
Mino et al.~\cite{mino} and Quinn and Wald~\cite{quinn}  proved that the regularized radiation reaction is solely contributed by $h_{\rm ret}$ which is everywhere continuous but not necessarily smoothly differentiable. This approach is useful when one knows the Green's function of the background spacetime. 

The second method, usually called Detweiler-Whiting decomposition, keeps the point-particle description of the problem, but instead separates the total metric perturabtion field $h$ into a regular piece $h_R$ and a singular piece $h_S$ \cite{detweiler}.  The singular piece diverges at the particles's location but does not have any effect on the particle's motion. It can be obtained by either transforming its expression in local THZ (Thorne-Hartle-Zhang) coordinate system \cite{detweiler2} to the background coordinate system or applying the Hadamard decomposition of the Green's function \cite{roland}. The regular field satisfies the homogeneous Einstein's equation and is responsible for the geodesic deviation of the particle's motion in background spacetime; it is obtained by subtracting the singular field from the full field. 


Regularization procedures above normally provides, in a particular gauge, a force in terms of a geodesic orbit of a particle.  In order to obtain the evolution of the particle and the out-going wave, one must construct an algorithm to compute the force, and use it to drive one's particle trajectory away from geodesic motion \cite{gralla}.     There are two major approaches towards the numerical implementation. One way is the {\it mode sum} approach, developed by Barack and Ori \cite{barack4}, which decomposes each of the 10 metric components into spherical harmonics, and solve 10 coupled 1+1 wave equations for each $(l,m)$. Because of the modal decomposition, metric component for each $(l,m)$ is finite even at the particle's location.  The particle equation of motion is then regularized mode-by-mode, by subtracting a series of regularization parameters for each $(l,m)$ --- these regularization parameters can be calculated  either from the singular field of Detweiler and Whiting, or to the direct part of Mino,  Sasaki and Tanaka. This mode sum method has already been implemented by Warburton et al.~\cite{warburton} for Schwarzschild gravitational EMRI problem. 

In the second approach, one directly applies a 3+1 decomposition of spacetime, and tries to obtain the regular field directly --- by obtaining a field $\tilde h_R$ which is approximately the Detweiler-Whiting $h_R$ near the particle, but gradually becomes the full field at null infinite and near horizon.  As shown by Vega and Detweiler \cite{Ian1}, the field $\tilde h_R$ satisfies a wave equation with out-going boudnary condition at infinity and horizon, but with a source that can be computed from the Detweiler-Whiting singular field $h_S$.   Diener and Vega \cite{diener} have implemented this method for a scalar particle orbiting a Schwarzschild black hole. In principal, this effective source method is also suitable for 1+1 evolution scheme. In practice, translating the 3+1 effective source into spherical decomposed form and implementing it into a working code still requires substantial amount of work.  

In this paper, we formulate a Hamiltonian approach towards the EMRI problem in Schwarzschild background, with the aim of providing a new angle to view this problem. We will only write down the equations, but not attempt to solve them numerically.  We start by generalizing Moncrief's (already spherical harmonic decomposed) quadratic action of perturbative Schwarzschild spacetime to include a point particle, and write down a joint Hamiltonian for the particle and the spherical harmonic decomposed  field. 
%
%
The total Hamiltonian leads naturally to a set of canonical equations that describe the joint evolution of the particle and the field.  
Moreover, since we are adopting Moncrief's formalism, gauge invariant part of the perturbation fields are separated out from the rest of the degrees of freedom --- these fields, together with lapse and shift, drive the rest of the fields. 
For each $(l,m)$, there are 6 pairs of canonical quantities; 2 pairs are always gauge invariant, and evolve independently (but driven by the particle); among the other 4 pairs, 3 canonical momenta and 1 canonical coordinate correspond to the momentum and Hamiltonian constraints, while the other 3 canonical coordinate and 1 canonical momentum can be fixed by gauge choices.  

Although the Hamiltonian approach provides a rather good way of organizing the fields, we have not found any stand-alone regularization technique --- and must instead adopt an existing one.  In principle, taking the 4-D Detweiler-Whiting singular field $h_S$ in any gauge, as long as their continuity survives the differential operations required for constructing our canonical field quantities, they can be readily used to obtain effective sources for $\tilde h_R$, the effective regular field.  However, the currently available singular field only allows the use of Lorenz gauge, which means we will have to fix that gauge, evolving the 8 above mentioned equations. 


This paper is organized as follows. in Sec.~\ref{sec2} we briefly review  Moncrief's Hamiltonian approach for gravitational perturbations of the Schwarzschild metric. After that we introduce additional terms into the action that describe the point particle. From this new action, we rederive the odd and even parity metric perturbation master equations as well as the point mass equations of motion in Sec.~\ref{sec3}. Note that for both odd and even parities, wave equations agree with the known master equations for Regge-Wheeler and Zerilli-Monrief functions with a point-particle source \cite{Karl, Karl2,Karl3}. On the other hand, the point mass equations of motion are now subject to the influence of background metric and both odd and even parity metric perturbations. They have the physical meaning of the geodesic motion in the perturbed background spacetime. In Sec.~\ref{sec4}, we will discuss possible ways to regularize the gauge invariant quantities and specific gauge choice, and hence obtain fully regularized set of equations for both the metric perturbations and the point mass. We conclude in Sec.~\ref{sec5}.

\section{Review of Moncrief's Hamiltonian Approach}\label{sec2}
The Arnowitt, Deser, and Misner (ADM) approach to general relativity \cite{adm} was established more than fifty years ago. In this approach, the Einstein-Hilbert action is written in a $3+1$ format  similar to a constrained Hamiltonian dynamical system: the  spatial 3-metric components are treated as canonical coordinates, while their conjugate momenta are related to components of the extrinsic curvature.
 The lapse and shift functions serve as Lagrange multipliers for the Hamiltonian and momentum constraints.  As one varies the action with respect to the canonical coordinates (not including lapse and shift functions) and their conjugate momenta, a set of evolution equations can be obtained.  As one varies the action with respect to the lapse and shift functions, a set of constraint equations are obtained --- these equations are to be satisfied at the initial time, and will keep being satisfied if the above-mentioned evolution equations are followed. 
This approach rewrites Einstein's equations as an initial-value problem; it is closely related to the modern development of numerical relativity \cite{shapiro}.

\subsection{First-order perturbation of a static space-time in 3+1 form}

Starting from this section, we review Moncrief's implementation of the ADM formalism to perturbed Schwarzschild spacetimes~\cite{moncrief}. 
In general, for a static background spacetime, if we take spatial slices orthogonal to the time-like Killing vector $\vec \partial_t$, and use integration curves of $\vec \partial_t$ to identify points with the same spatial coordinates (on the spatial slices), we will have a constant 3-metric $\gamma_{ij}$ (with determinant $\gamma$), vanishing extrinsic curvature, a lapse $N$ that only depends on spatial coordinates, and a vanishing shift vector $N_j=0$. Here and henceforth in the paper, we shall use $i,j,k,...=1,2,3$ to label spatial coordinates on each slice.  If we have a perturbed 3-metric $h_{ij}$, their canonical conjugates $p^{ij}$, lapse perturbation $N'$, and shift perturbation $N'_j$, then the perturbative part of Einstein-Hilbert action, up to quadratic order in these perturbative quantities, can be written as: \cite{moncrief}
\begin{equation}\label{eqnomass}
J = \int d^4 x \left [ p^{ij} \frac{\partial h_{ij}}{\partial t}  -N'_i {\mathcal{H}'}^{i}-N'\mathcal{H}' - N \mathcal{H}^*\right ]
\end{equation}
Here $\mathcal{H}'$ and ${\mathcal{H}'}^i$ are the Hamiltonian and momentum constraints, respectively, 
\begin{subequations}
\begin{align}
{\mathcal{H}'}^{i} =& -2p^{ij}_{|j} \\
\mathcal{H}' =& -\gamma^{1/2}\left [ {h_{ij}}^{|ij} -{h_{|i}}^{|i}-{h_{ij}}^{(3)}R^{ij}\right ]
\end{align} 
\end{subequations}
and
\begin{eqnarray}
\mathcal{H}^* &= \gamma^{-1/2}\left [ p^{ij}p_{ij}-\frac{1}{2}p^2\right ]+\frac{1}{2}\gamma^{1/2}  \frac{1}{2}h_{ij|k}h^{ij|k}   \nonumber \\
& +\frac{1}{2} \gamma^{1/2} \left [  -h_{ij|k}h^{ik|j}-\frac{1}{2}h_{|i}h^{|i}+2h_{|i}{h^{ij}}_{|j}\right ] \nonumber \\
&+\frac{1}{2} \gamma^{1/2} \left [h{h^{ij}}_{|ij} -h {h_{ij}}^{(3)}R^{ij} \right ]
\end{eqnarray}
Here $^{(3)}R_{ij}$ is the Ricci tensor associated with $\gamma_{ij}$. The covariant derivative ``$|$'' here is with respect to the background $3$ metric $\gamma_{ij}$.



The action $J$ in Eq.~(\ref{eqnomass}) leads to a Hamiltonian system with constraints.  In particular, variation with respect to the lapse function $N'$ and shift function $N'_j$ gives rise to the constraint equations,
\begin{equation}
{\mathcal{H}'}^{i} =0,\ \mathcal{H}' = 0
\end{equation}
while variations with respect to $h_{ij}$ and $\pi^{ij}$ gives rise to the evolution equations:
\begin{equation}
\frac{\partial h_{ij}}{\partial t} =\frac{\delta H_T}{\delta \pi^{ij}},\  \frac{\partial \pi^{ij}}{\partial t} = -\frac{\delta H_T}{\delta h_{ij}}
\end{equation}
Here we have defined 
\begin{equation}\label{ham}
H_T = \int d^3 x \left[ N \mathcal{H}^*+N' \mathcal{H}' +N'_i {\mathcal{H}'}^{i} \right ]
\end{equation}

\subsection{Degrees of freedom} 
\label{subsec:dof}

Let us now count the number of degrees of freedom of this Hamiltonian system. Nominally, we start from 6 metric perturbations, plus 6 canonical momenta, governed by 6 pairs (i.e., 12) equations of motion.  However, we have 4 constraints and 4 gauge degrees of freedom at all times; so in principle we should be able to cut down to 4 gauge independent functions, or 2 pairs of canonical degrees of freedom --- this is what Moncrief worked out explicitly for Schwarzschild.   

From a 3+1 point of view, we need to show that we indeed only have 4 independent data to specify for free at the initial time slice --- and the evolution of these 4 independent data can already describe all solutions.   For this, we note that when specifying the 12 {\it initial} perturbation functions, we need to subject them to 4 constraints, so there are 8 independent remaining degrees of freedom.  However, we have 3-D gauge within the slice, as well as an extra parameter determining the slicing, so we have 4 additional functions that can be used to reveal initial data that are actually equivalent to each other --- with 4 truly independent degrees of freedom left.  The {\it evolution} these 4 functions will be supplemented by the 4 constraints and the 4 lapse/shift functions to complete the 12 functions. 

In fact, we can make the above arguments a little more precise.  Suppose, after a canonical transformation, we can transform the Hamiltonian and momentum constraints to become independent canonical coordinates and momenta.  More specifically, let us label the Hamiltonian constraint the first canonical coordinate $\mathcal{Q}_0$, denote the conjugate momentum of $\mathcal{Q}_0$ as $\mathcal{P}_0$; let us then label the momentum constraints as $\mathcal{P}_{1,2,3}$, and label their conjugate coordinates as $\mathcal{Q}_{1,2,3}$. In other words, we have
\begin{equation}
\label{canonicalconstraints}
\mathcal{Q}_0 =\mathcal{H}' \,,\quad \mathcal{P}_i = \mathcal{H}'^i \,,\quad i=1,2,3.
\end{equation}
We will label the rest of the canonical coordinates $\mathcal{Q}_{4,5}$ and momenta $\mathcal{P}_{4,5}$.  Noting that all components of the momentum constraints already have vanishing Poisson brackets with each other, 
\begin{equation}
\big\{\mathcal{H'}^i,\mathcal{H'}^j\big\}=0\,,\quad i,j=1,2,3.
\end{equation}
we only need to make sure that the Hamiltonian constraint have a vanishing Poisson bracket with all components of the momentum constraint:
\begin{equation}
\label{commuteHHj}
\big\{ \mathcal{H'},\mathcal{H'}^j \big\}=0\,. \quad j=1,2,3.
\end{equation}
This rather straightforward to show if we look at the evolution equation for the Hamiltonian constraint:
\begin{equation}
\label{Hconst}
\frac{d}{dt} \mathcal{H'}  = \big\{\mathcal{H'},\mathcal{H'}^j\big\} N_j' +\mbox{(terms not involving shift)}\,.
\end{equation}
Now imagine we already have vanishing constraints initially, then in order to guarantee vanishing constraints during the subsequent evolution --- regardless of the shift function $N_j'$,  we must ensure that the Poisson bracket $\big\{\mathcal{H'},\mathcal{H'}^j\big\}$ vanish numerically.  However, for linear perturbation theory, $\mathcal{H'}$ and $\mathcal{H'}^j$ are linear in the canonical coordinates and momenta, $h_{ij}$ and $p^{ij}$, their Poisson brackets are simply numbers (or rather, functions of the spatial coordinate) that do not depend on these perturbative fields.  In this way, the numerical vanishment of $\big\{\mathcal{H'},\mathcal{H'}^j\big\}$ in Eq.~\eqref{Hconst} is equivalent to Eq.~\eqref{commuteHHj} --- hence Eq.~\eqref{canonicalconstraints} is always possible.

Next, let us consider the consequence of the important property that once $\mathcal{Q}_0$ and $\mathcal{P}_{1,2,3}$ starts from 0, they must keep being 0.  This means their time derivatives must only contain themselves --- which means, in the Hamiltonian, quantities $\mathcal{P}_0$ and $\mathcal{Q}_{1,2,3}$ must only multiply $Q_{0}$ and $\mathcal{P}_{1,2,3}$
\begin{equation}
  \mathcal{P}_0 \cdot\left[\mbox{only $Q_0$ and $\mathcal{P}_{1,2,3}$}\right] \,\&\, \mathcal{Q}_{1,2,3}\cdot \left[\mbox{only $Q_0$ and $\mathcal{P}_{1,2,3}$}\right]
\end{equation}
The absence of $\mathcal{Q}_{4,5}$ and $\mathcal{P}_{4,5}$ in the above terms means that the derivative of $\mathcal{P}_{4,5}$ and $\mathcal{Q}_{4,5}$ cannot include $\mathcal{P}_0$ or $\mathcal{Q}_{1,2,3}$.  This, plus the fact that $\mathcal{Q}_0$ and $\mathcal{P}_{1,2,3}$ vanishes, means that the evolution of $\mathcal{P}_{4,5}$ and $\mathcal{Q}_{4,5}$ must be self contained, or:
\begin{equation}
\frac{d}{dt}{\mathcal{{P}}}_{4,5} \sim \mathcal{P}_{4,5} \& \mathcal{Q}_{4,5}\,,\quad
\frac{d}{dt}{\mathcal{{Q}}}_{4,5} \sim \mathcal{P}_{4,5} \& \mathcal{Q}_{4,5}\,.
\end{equation}
In this way, these 4 are the gauge-invariant canonical variables. Another 4 equations are driven by the above gauge-invariant quantities, plus lapse and shift:
\begin{eqnarray}
\frac{d}{dt}{\mathcal{{P}}}_{0} &\sim &\mathcal{P}_{4,5} \, \& \, \mathcal{Q}_{4,5}\, \&\,  \mathcal{P}_0 \, \& \, \mathcal{Q}_{1,2,3}\& N' \\
\frac{d}{dt}{\mathcal{{Q}}}_{1,2,3} &\sim& \mathcal{P}_{4,5} \, \& \, \mathcal{Q}_{4,5}\, \&\,  \mathcal{P}_0 \, \& \, \mathcal{Q}_{1,2,3}\& N'_{1,2,3}\,.
\end{eqnarray}
The final 4 equations are simply that the constraints vanish.

As we shall see below, in his treatment of Schwarzschild perturbations, Moncrief did follow the above general prescription --- by directly using constraints as canonical coordinate and momenta.   Note that this structure seems rather generic, and does not seem to be limited to Schwarzschild or even static spacetimes --- of course, it is another issue whether one can separate these into different $(l,m)$ components. 

\begin{table}
\begin{tabular}{c|c|c}
& Odd Parity & Even Parity  \\
\hline
Lapse & & $H_0$ \\
Shift  & $h_0$ ($l \ge 1$) & $H_1$, $h_0^*$ ($l\ge 1$) \\
3-Metric &  $h_1$ ($l \ge 1$),  $h_2$ ($l\ge 2$) & $H_2$, $K$,  $h_1^*$ ($l\ge 1$) ,  $G$ ($l\ge 2$)
\end{tabular}
\caption{List of lapse, shift and 3-metric perturbations \label{tab:fields}}
\end{table}

\subsection{Schwarzschild Perturbations}

Let us return to perturbations of Schwarzschild.  In our case, the background metric is 
\begin{equation}
ds^2 = -\left(1-\frac{2M}{r}\right)dt^2+\frac{dr^2}{1-\frac{2M}{r}}+r^2(d\theta^2+\sin^2{\theta} d\phi^2)
\end{equation}
and we choose to start with constant-$t$ slices in this Schwarzschild coordinate system, and adopt spatial coordinates 
\begin{equation}
(x^1,x^2,x^3)=(r,\theta,\phi). 
\end{equation}
In this way, we have $N=\sqrt{-1/g^{00}}=\sqrt{1-2M/r}$, and non-zero components of $\gamma_{ij}$ given by
\begin{equation}
\gamma_{11} = \frac{1}{1-2M/r}\,,\quad
\gamma_{22}= r^2\,,\quad
\gamma_{33}=r^2\sin^2\theta\,.
\end{equation}
We shall use $i,j,k=1,2,3$ to label these spatial coordinates, and  write components of the metric perturbation $h_{ij}$ as functions of spacetime coordinates $(t,r,\theta,\phi)$ ---  and separate the angular dependence by decomposing them into scalar, vectorial and tensorial spherical harmonics \cite{Karl}
\begin{subequations}
\label{eqg1}
\begin{align}
h^{lm}_{AB} & = r^2 \left[ K(t,r) U^{lm}_{AB}+G(t,r) V^{lm}_{AB}\right ]+h_2(t,r)W^{lm}_{AB}  \\
h^{lm}_{rr}&=1/f H_2(t,r)Y^{lm} \\
h_{rA} &= h^*_1(t,r)Z^{lm}_A+h_1(t,r)X^{lm}_A
\end{align}
\end{subequations}
Here we have defined $f \equiv 1-2M/r$, and shall use $A,B,...=1,2$ to label angular coordinates 
\begin{equation}
(\Omega^1,\Omega^2) = (\theta, \phi). 
\end{equation}
The conjugate momenta $p_{ij}$ can be similarly decomposed, while the decomposition of lapse and shift perturbations $N', N'_i$ are \cite{Karl}
\begin{subequations}
\label{eqng2}
\begin{align}
h_{tt}^{lm} =& f H_0(t,r) Y^{lm} \\
 h_{tr}^{lm}=&H_1(t,r)Y^{lm}  \\
h_{tA}^{lm} =& h^*_0(t,r) Z^{lm}_A+h_0(t,r)X_A^{lm}
\end{align}
\end{subequations}
Here we have used odd parity vector and tensor spherical harmonics $X^{lm}_A, W^{lm}_{AB}$, as well as even parity ones, $Z^{lm}_A, U^{lm}_{AB},V^{lm}_{AB}$; their definitions can be found in \cite{Karl, Karl2}; we have also listed them in Appendix \ref{ap1}. 
For each $(l,m)$, ($l\ge 2)$,  we have a total of $10$ independent functions characterizing $10$ independent metric components; $4$ of them are lapse and shift perturbations: $H_0$ is lapse perturbation and $H_1,h^*_0,h_0$ are shift perturbations. The rest 6 functions are spatial metric perturbations: $K,G,h_2,H_2,h^*_1,h_1$.  For even parity, there are  $1$ lapse perturbation function $H_0$, $2$ shift perturbation functions $H_1,h^*_0$ and $4$ spatial metric perturbation function $K,G,H_2,h^*_1$. For odd parity, there are no lapse perturbation, $1$ shift perturbation function $h_0$ and $2$ spatial metric perturbation functions $h_1,h_2$ for odd parity. This counting is shown in Table~\ref{tab:fields}.

\subsection{Odd Parity ($l\ge 2$)}

Let us first look at odd-parity perturbations, which contain spatial-metric perturbations $h_1,\,h_2$ [Eq.~\eqref{eqg1}] and shift perturbation $h_0$ [Eq.~\eqref{eqng2}]. 
%
%
All odd-parity infinitesimal coordinate transformations within the spatial slice can be represented using the odd-parity vector harmonic $X^{lm}_A$, 
\begin{equation}
\label{eqocoord}
\Omega'_A = \Omega_A+\sum_{lm}C^{lm}(r,t)X^{lm}_A =\Omega_A+C_A \,,\\
\end{equation}
which, after applying 
\begin{equation}
\delta h_{ij} =C_{i|j}+C_{j|i}
\end{equation}
lead to 
\begin{align}
\delta h_1=C_{,r}-\frac{2}{r}C, \quad \delta h_2 = -2C\,.
\end{align} 
Moncrief defined new perturbation functions 
\begin{equation}
k_1 = h_1+\frac{1}{2}\left(h_{2,r}-\frac{2}{r} h_2\right)\,,\quad  k_2=h_2
\end{equation}
which transform as
\begin{align}
\delta k_1 =0, \, \delta k_2 = -2C
\end{align}
%
%
%
In other words, $k_1$ is invariant under infinitesimal coordinate transformations while $k_2$ is vulnerable to the specific choice of gauge.  In terms of $k_{1,2}$, and their canonical conjugates $\tau_{1,2}$, the 
the odd-parity Hamiltonian [Eq.~(\ref{ham})]  can now be expressed as
\begin{eqnarray}
H_T &=& \frac{1}{\lambda+1}\int dr \left\{ \tau^2_1+\frac{r^2 f}{\lambda}\left [ \tau_2-\frac{1}{2}\tau_{1,r}-\frac{1}{r}\tau_1 \right ]\right\}  \nonumber \\
&+& 2\lambda(\lambda+1) \int dr \ \frac{f}{r^2} k^2_1 -2 \int dr \  h_0 \tau_2
\end{eqnarray}
with 
\begin{equation}
\lambda \equiv (l-1)(l+2)/2
\end{equation}
Variation of the shift function $h_0$ in the Hamiltonian gives the odd-parity momentum constraint equation 
\begin{equation}
\tau_2=0
\end{equation}

Equations of motion for the dynamical variables take the form \cite{moncrief}
\begin{subequations}
\label{eqmo1}
\begin{align}
\frac{\partial k_1}{\partial t}  = &\frac{\delta H_T}{\delta \tau_1} 
=\frac{\tau_1}{2(\lambda+1)}
+\frac{\displaystyle r^2 \left [  f \left (\tau_2-\frac{(r^2 \tau_1)_{,r}}{2r^2}\right) \right]_{,r}}{2\lambda(\lambda+1)} \\
\frac{\partial \tau_1}{\partial t} = & -\frac{\delta H_T}{\delta k_1} = -\frac{4\lambda(\lambda+1)}{r^2}f k_1  \\
\frac{\partial k_2}{\partial t}  = &\frac{\delta H_T}{\delta \tau_2} = \frac{f r^2}{\lambda(\lambda+1)}\left [\tau_2-\frac{{(r^2 \tau_1)_{,r}}}{2r^2}
\right ]-2h_0  \\
\frac{\partial \tau_2}{\partial t}  = &-\frac{\delta H_T}{\delta k_2}=0
\end{align}
\end{subequations}
Here $\tau_2=0$ is constraint; $(k_1,\tau_1)$ is the gauge-invariant sector, which evolves independently (once setting $\tau_2=0$).  Gauge is fixed by choosing $h_0$, which correspondingly fixes the evolution of $k_2$. (Note that $\tau_2$ is constraint and should vanish.)  As an example, the Regge-Wheeler gauge is obtained by imposing that $k_2=0$, which requires setting 
\begin{align}
h_0 = \frac{f(r^2\tau_1)_{,r}}{4\lambda(\lambda+1)}
\end{align}
%
These odd-parity perturbation equations determine $2$ out of the $6$ spatial 3-metric components, and 3 out of the total $10$ spacetime 4-metric components.

The Regge-Wheeler function frequently used in the literature to describe odd-parity perturbations~\cite{Karl, Karl2, Karl3, carlos, carlos2, carlos3} is related to  $k_1$ by 
\begin{equation}
\psi_{\rm RW} = f k_1/r
\end{equation}
$\psi_{\rm RW}$ is invariant under infinitesimal gauge transformations.

\subsection{Even Parity  ($l\ge 2$)}
For even parity, there are 4 spatial-metric perturbations, $K, G, H_2, h^*_1$  [Eq.~\eqref{eqg1}], 1 lapse perturbation $H_0$ and 2 shift perturbations $H_1, h^*_0$ [Eq.~\eqref{eqng2}]. Moncrief found it convenient to recombine $K, G, H_2, h^*_1$ and define a new set of variables $q_1, q_2, q_3, q_4$. Like $k_1$ for odd parity perturbation, $q_1$ is invariant under infinitesimal gauge transformation whereas $q_2,q_3,q_4$ are gauge dependent. The conversion between $K, G, H_2, h^*_1$ and $q_1, q_2, q_3, q_4$ can be found in \cite{moncrief} and Appendix \ref{ap2} in this paper. In terms of the new coordinates and their conjugate momenta, $\pi_{1,2,3,4}$, the even parity Hamiltonian is given by:
\begin{eqnarray}
H_T &=& \int dr \left \{-\frac{2}{r}f\pi_4 \left [ r(\pi_1-\pi_{2,r})+\left(1-\frac{M}{f r}\right)\pi_2 \right ] \right \} \nonumber \\
&+& \int dr \left \{ \frac{f }{2 r^2 \lambda}\left [ \frac{ \pi^2_3}{\lambda+1}+2\pi_3\left [ r \Lambda \pi_1+\pi_2(r \Lambda)_{,r}\right ]\right ] \right \}\nonumber \\
&+& \int dr \left \{  \frac{\lambda+1}{2\lambda}f\Lambda^2\pi^2_1 + \frac{\pi^2_4}{4(\lambda+1)} -\frac{\lambda}{r \Lambda^2}q_1q_2 \right \} \nonumber \\
&+& \int dr \left \{ \frac{\lambda f}{2 (\lambda+1)\Lambda^2}(q_2-q_{1,r})^2 +\frac{2 \lambda^2}{r^2 \Lambda^3} q^2_1 \right \}\nonumber \\
&-& \int dr \left \{ \frac{M q_2(q_2-q_{1,r})}{2 (\lambda+1)\Lambda r}+\frac{M q_2}{2 r}(r q_{3,r}-\frac{2}{r}q_4)\right \} \nonumber \\
&+& \int dr \left \{ \frac{H_0 q_2}{2}+H_1 \pi_4 +h^*_0 \left [ \frac{2 \pi_3}{r^2} -\frac{(r^2 \pi_4)_{,r}}{r^2}\right ]\right \}. 
\end{eqnarray}
Here $\Lambda$ is defined as 
\begin{equation}
\Lambda\equiv  2\left(\lambda+\frac{3M}{r}\right) =(l-1)(l+2)+\frac{6M}{r}\,.
\end{equation}
From this Hamiltonian, it is straight forward to obtain the following canonical equations of motion:
\begin{subequations}
\label{eqe}
\begin{align}
\dot{q_1} =& -2 f \pi_4 +\frac{f \Lambda}{\lambda r}\pi_3+\frac{\lambda+1}{\lambda}f \Lambda^2 \pi_1  \\
\dot{\pi_1} =& \frac{\lambda q_2}{r \Lambda^2}-\frac{4 \lambda^2}{r^2\Lambda^3}q_1-\left [ \frac{\lambda f}{(\lambda+1)\Lambda^2}(q_2-q_{1,r}) \right ] _{,r}\nonumber \\
+&\left [ \frac{M q_2}{2(\lambda+1)\Lambda r} \right ]_{,r}  \\
\dot {q_2} =& \frac{2 f}{r^2}\pi_3-\frac{2}{r}\left(f-\frac{M}{r}\right)\pi_4-2 (f \pi_4)_{,r} 
 \\
\dot{\pi_2} =& \frac{\lambda}{r \Lambda^2}q_1-\frac{\lambda f}{(\lambda+1)\Lambda^2}(q_2-q_{1,r})-\frac{M}{2 r}\left(r q_{3,r}-\frac{2}{r}q_4\right) \nonumber \\
+& \frac{M(2 q_2-q_{1,r})}{2(\lambda+1)\Lambda r}-\frac{H_0}{2}  
\label{pi2dot}
\\
\dot{q_3} =& \frac{f \pi_3}{r^2 \lambda(\lambda+1)}+\frac{f }{r^2 \lambda}\left[ r \Lambda \pi_1+ 2 \lambda \pi_2 \right ]+\frac{2 h^*_0}{r^2}  
\label{q3dot}
\\
\dot{\pi_3} =& -\frac{M}{2}q_{2,r}  \\
\dot{q_4} =& \frac{\pi_4}{2(\lambda+1)}+H_1+r^2\left ( \frac{h^*_0}{r^2}\right )_{,r} \nonumber \\
-& 2 f(\pi_1-\pi_{2,r})-\frac{2}{r}\left ( f-\frac{M}{r}\right )\pi_2  
\label{q4dot}
\\
\dot{\pi_4} =& -\frac{M}{r^2}q_2
\end{align} 
\end{subequations}
By varying the lapse perturbation $H_0$ and the shift perturbations $H_1, h^*_0$, it is straightforward to obtain the hamiltonian constraint equation 
\begin{equation}
q_2=0
\end{equation}
 as well as the even-parity momentum constraint equations 
 \begin{equation}
 \pi_3 =\pi_4 =0.
 \end{equation} 
Note that $(q_1,\tau_1)$ is the even-parity gauge-invariant sector; $(\pi_2,q_3,q_4)$ are the gauge-dependent sector, which are determined after the lapse $H_0$ and shifts $(H_1,h^*)$ are fixed. 

For example, the Even-Parity Regge-Wheeler gauge is described by $q_3= q_4=\pi_2=0$, which requires initially setting $q_3=q_4=\pi_2=0$, and keeping it true by imposing $\dot{q}_3=\dot{q}_4=\dot{\pi}_2=0$ through setting the appropriate $h_0^*$ [Eq.~\eqref{q3dot}], $H_1$ [Eq.~\eqref{q4dot}] and $H_0$ [Eq.~\eqref{pi2dot}].  These even-parity perturbation equations determine the other 4 of the 6 spatial 3-metric perturbations, and the other 7 out of the the 10 spacetime 4-metric perturbations.

Gauge invariant quantity $\psi_{\rm ZM}$ are commonly \cite{Karl, Karl2, Karl3, carlos, carlos2, carlos3}  used for even parity perturbation and it is defined by $q_1/(\lambda+1)/\Lambda$ or equivalently \cite{carlos2}
\begin{equation}
\psi_{\rm ZM}=\frac{r}{\lambda+1}\left[ K+\frac{2 f}{\Lambda}\left(H_2-r\frac{\partial K}{\partial r}\right)\right ] +\frac{2 f }{\Lambda}\left [\frac{r^2 \partial G}{\partial r}-2 h^*_1 \right ]
\end{equation}

\subsection{Monopole and dipole perturbations}

For $l\le 1$, the evolution of  3-metric perturbations can all be fixed by the constraint equations plus arbitrary choices of lapse and shift functions.   More specifically:

For $l=0$, there are only even-parity perturbations.  We have lapse perturbation $H_0$ and shift perturbation $H_1$, plus metric perturbations $H_2$ and $K$ --- while lapse perturbation $h_0*$ and 3-metric perturbations $G$ and $h_1^*$ all vanish due to the non-existence of the vector and tensor harmonics $Z$ and $V$.   However, there still exists one Hamiltonian constraint and one momentum constraint.  We can transform $(H_2,K)$ into two new canonical coordinates, one of them the Hamiltonian constraint, the other the canonical conjugate of the momentum constraint --- leaving no gauge-invariant perturbation fields.

For $l=1$, even parity perturbation, we have lapse perturbation $H_0$, shift perturbations $H_1$ and $h_0^*$, plus three non-vanishing 3- metric perturbation fields, namely $H_2$, $K$ and $h_1^*$.  However, there exists 1 Hamiltonian constraint and two momentum constraints, and we can transform $(H_2,K,h_1^*)$ into the Hamiltonian constraint and the canonical conjugate of the two momentum constraints, also leaving no gauge-invariant perturbation fields.

For $l=1$, odd parity perturbation, we shift perturbation $h_0$, and one metric perturbation field, which is $h_1$.  We also have one momentum constraint, therefore it means a spatial operation on $h_1$ will become the canonical conjugate of the momentum constraint, this means we have no gauge-invariant perturbation field as well.

\section{3+1 approach with point mass source}\label{sec3}
 In this paper, we are interested in the joint evolution metric perturbations and the motion of a point particle.  In this section,  we will augment Moncrief's formalism with a point particle.
 
 \subsection{3+1 Formulation}
 
 Since Eq.~(\ref{eqnomass}) is the action for free metric perturbations alone, we need to add the action for the point particle. Using the prescriptions in \cite{adm}, we can write:
 \begin{eqnarray}
 J_m &=& m\int d\tau  \nonumber \\
 &=& \int d^4 x \ \delta^{(3)}({\bf r}-{\bf Q}(t))\left [ P_i \frac{\partial x^i}{\partial t} -\mathcal{N}\left ( g^{ij}P_iP_j+m^2\right )^{1/2} \right ] \nonumber \\
 &+&\left [ \mathcal{N}^iP_i\delta^{(3)}({\bf r}-{\bf Q}(t))\right ]
  \end{eqnarray}
Here, regarding quantities of space-time geometry, we have $g_{ij} = \gamma_{ij}+h_{ij}$ the total perturbed spatial metric; $\mathcal{N}=N+N'$ the total lapse, and $\mathcal{N}_i=N'_i$ the total shift (recall that $N_i =0$); regarding the particle, $P_j$ ($j$=1,2,3) are components of the 3-momentum, and $\mathbf{Q}(t)$ represents the spatial coordinates of the particle, which are, more specifically, $(R(t),\Theta(t),\Phi(t))$. The $\delta$ function is more explicitly written as
\begin{equation}
\delta^{(3)}(\mathbf{r} -\mathbf{Q}(t)) = \delta (r-R(t)) \delta (\theta-\Theta(t)) \delta (\phi-\Phi(t))
\end{equation}

From this action, we can read off the part of the Hamiltonian that involves the point particle, which includes the Hamiltonian of the point particle alone
\begin{equation}
H_{m}(Q^k,P_k) = N(Q^k) \sqrt{\gamma^{ij}(Q^k)P_iP_j+m^2}
\end{equation}
which describes the geodesic motion of the particle,  plus the interaction Hamiltonian that couples the particle and metric perturbations,
\begin{align}\label{eqhadd}
& H_{\rm int}(Q^k,P_k,N',N_j',h_{ij}) \nonumber\\
=& N'\sqrt{\gamma^{ij} P_iP_j+m^2}
-\frac{N}{2}\frac{h_{ij} \gamma^{il} \gamma^{jm} P_l P_m}{ \sqrt{\gamma^{ij} P_iP_j+m^2}} 
\nonumber\\
-& N'_jP^j
\end{align}
Here we have suppressed $N'$, $N$, $N_j'$, $\gamma^{ij}$ and $h_{ij}$'s dependence on $Q^k$, for simplicity --- but the reader is reminded that $Q^k$ enters this interaction Hamiltonian through these quantities' dependence on $Q^k$. Note that $H_{\rm int}$ is {\it linear} in the metric perturbations $N'$, $N_j'$ and $h_{ij}$.


The total hamiltonian for the combined system of point particle plus metric perturbations is 
\begin{equation}
H_{\rm tot}  = \sum_{lm}(H^{lm}_{\rm oddT}+H^{lm}_{\rm evenT})+16 \pi H_{\rm m} + 16\pi H_{\rm int}
\end{equation}
The $16 \pi $ is actually the $2 \kappa=2(8\pi G)$ factor in the Einstein-Hilbert action and we are taking the Newton's constant $G$ to be unity.  We have now enlarged the set of canonical coordinates and momenta to include $(Q^k,P_k)$.  
%
%
%
The field-alone term in $H_{\rm tot}$ describes the free propagation of metric perturbations around Schwarzschild,  the $H_m$ term describes the geodesic motion of the point particle;  it is $H_{\rm int}$ that couples the fields and the particle together: it allows the particle's motion to drive field perturbations, and field perturbations to act back onto the particle --- with action and back-action described in a self-consistent way.

Due to spherical symmetry of the background spacetime, we can assume that the point particle is confined within the equatorial plane, with $\Theta=\pi/2$, $P_\theta=0$ --- and we only need to deal with  $R(t),\Phi(t)$.  In addition, because $H_{\rm int}$ is linear in metric perturbations, we can divide it into sum of an odd-parity component and an even-parity component --- each component only involving one type of metric perturbations.

\subsection{Odd Parity $(l \ge 2)$}

Odd-Parity metric perturbations are described in terms of odd-parity vector and tensor harmonics $X^{lm}_A,W^{lm}_{AB}$ in Eq.~(\ref{eqg1}) and Eq.~(\ref{eqng2}). For later convenience, we  denote by $\Pi$ and $\Xi$,
\begin{equation}
\label{eqpixi}
\Pi = W^{lm}_{AB} P^AP^B, \ \Xi = X^{lm}_A P^A
\end{equation}
the contractions of these harmonics with angular components of momentum.  Plugging Odd-Parity perturbations in Eqs.~\eqref{eqg1} and \eqref{eqng2} into Eq.~\eqref{eqhadd}, taking Eq.~\eqref{eqpixi} into account, we obtain, for each $(l,m)$,  
%
\begin{eqnarray}
\label{Hintodd}
&&H_{\rm int}^{\rm odd} \nonumber\\
&=& \left. \left [ -h_0\Xi -\frac{\sqrt{f}}{2}\frac{2 h_1P^r\Xi+h_2\Pi}{\sqrt{\gamma_{ij}P^iP^j+m^2}}\right ]\right |_{r=R(t),\,{{\Omega}}={{\tilde \Omega}}(t)} \nonumber\\
&=&\int dr \int d^2\Omega \ \delta(r-R(t))\delta^{(2)}(\Omega-\tilde\Omega(t)) \nonumber \\
&&\quad\left [ -h_0\Xi -\frac{\sqrt{f}}{2}\frac{2 (k_1-\frac{1}{2}k_{2,r}+\frac{1}{r}k_2)P^r\Xi+k_2\Pi}{\sqrt{\gamma_{ij}P^iP^j+m^2}}\right ].\quad 
\end{eqnarray}
Here we have used $\Omega$ to represent $(\theta,\phi)$, and $\tilde\Omega(t)$ to represent $(\Theta(t),\Phi(t))$.

For any expression inside the definition of $H_{\rm int}^{\rm odd}$, for example $\gamma_{ij}$,  it always appears along with a $\delta(r-R(t))$ function, therefore being a function of $(r,\theta,\phi)$ instead of $(R,\Theta,\Phi)$ does  not seem to make a difference.   However, because derivatives of fields are involved, we will encounter derivatives of $\delta$ functions in further calculations, and  for a generic function $\mathcal{G}(r)$
\begin{equation}
\mathcal{G}(r)\delta'(r-R(t)) \neq \mathcal{G}(R(t))\delta'(r-R(t))\,.
\end{equation}
This does not indicate an ambiguity in the equations of motion that we are ultimately going to obtain, but create intermediate steps that may differ. 
This requires us to be  careful with our conventions.   Here we shall use the convention that all terms in the integrand on the right-hand side of Eq.~\eqref{Hintodd}, apart from the $\delta$ functions, only depend explicitly on $(r,\theta,\phi)$, not on $(R,\Theta,\Phi)$.

Taking the above convention for Eq.~\eqref{Hintodd}, the new constraint equation is 
\begin{equation}\label{eqocon}
\tau_2 = -\frac{1}{2} \,  \Xi \,\delta(r-R(t))
\end{equation}
the right-hand side is singular at the  location of the point particle but zero elsewhere. Because $H_{\rm int}$ only contains metric perturbations, not their conjugate momenta, only the evolution of the momenta are affected.  The evolution equations for $\tau_1$ gains an additional term of:
\begin{equation}
\left.\frac{\partial \tau_1}{\partial t}\right|_{\rm add}= -16\pi \frac{\delta H_{\rm int}}{\delta k_1} = 16 \pi \,\Xi \frac{P^r}{P^0} \delta(r-R(t))
\end{equation}
%
 Similarly for $\tau_2$, the additional term is
\begin{eqnarray}
\left.\frac{\partial \tau_2}{\partial t}\right|_{\rm add} &=& -16 \pi \,\frac{\delta H_{\rm int}}{\delta k_2} = \frac{16\pi}{P^0}\left (\frac{\Pi}{2}-\frac{ P^r}{r}\Xi \right )\delta(r-R(t)) \nonumber \\
&+&\left [ \frac{8\pi P^r}{ P^0}\,\Xi\,\delta(r-R(t))\right ]_{,r} \nonumber \\
&=& \frac{16\pi}{P^0}\frac{\Pi}{2}\delta(r-R(t))-\frac{1}{r}\left. \frac{\partial \tau_1}{\partial t}\right |_{\rm add} -\frac{1}{2}\left[ \frac{\partial \tau_1}{\partial t}\right ]_{\rm add,r} \nonumber \\
\end{eqnarray}
It is easy to check that, up to linear order (i.e., inserting background geodesic equations of motion for the particle) this equation is consistent with the new constraint equation Eq~(\ref{eqocon}). Combining the evolution equations for $k_1, k_2, \tau_1, \tau_2$ and note $\psi_{\rm RW} = f k_1/r$, we can derive the following master equation for $\psi_{\rm RW}$
\begin{eqnarray}\label{eqol}
\left [ -\frac{\partial^2}{\partial t^2}+ \frac{\partial^2}{\partial r^{*2}}-V_{\rm odd}^l(r) \right ] \psi_{\rm RW}(r,t) = S_{\rm odd}(r,t)
\end{eqnarray}
where  we have defined
\begin{equation}
r^* = r+2M \log{(r/2M-1)}
\end{equation}
and
\begin{equation}
V_{\rm odd}^l=\frac{2f}{r^2}\left(\lambda+1-\frac{3M}{r}\right) \,,
\end{equation}
 The source term is
\begin{equation}
S_{\rm odd} = \frac{4\pi f}{(\lambda+1)r}\left [\frac{r^2}{\lambda}\left[ \frac{f \Pi \delta(r-R(t))}{P^0} \right]_{,r}- \frac{2\,P^r \Xi\delta(r-R(t))}{P^0}\right ]
\end{equation}
Here the subscript ``${\rm odd}$'' means odd parity. This source term agrees with the ones derived in literature \cite{zerilli, Karl, Karl2} as expected.

On the other hand, $H_{\rm int}^{\rm odd}$ introduces the following additional terms to the rate of change of the point particle's coordinates:
\begin{subequations}
\begin{align}\label{eqo1}
\left. \frac{d R}{d t}\right |_{\rm odd} = 16\pi \frac{\partial H_{\rm int}^{\rm odd}}{\partial P_r} 
=& -16\pi h_1\frac{f \, \Xi}{P^0}\left [ 1-\frac{(P_r)^2}{(P^0)^2}\right ] \nonumber\\&+ 8 \pi \frac{h_2 \Pi \,P_r}{(P^0)^3} \\
\left. \frac{d \Phi}{d t}\right |_{\rm odd} = 16\pi \frac{\partial H^{\rm odd}_{\rm int}}{\partial P_{\phi}}  
=& -16 \pi \left[ h_0 X^{\phi}-\frac{P_{\phi}[2h_1P^r\Xi+h_2\Pi]}{2 f r^2 \sin{\theta}(P^0)^3}\right] \nonumber \\
-& \frac{16 \pi}{P^0}\left [ h_1P^r X^{\phi}+h_2 W^{\phi \phi}P_{\phi}\right ]
\end{align}
\end{subequations}
where components of $X^{A}, W^{AB}$ can be found in Appendix \ref{ap1}. Similarly, the rate of change of the point particle's momenta also gains the following additional terms:
\begin{subequations}
\begin{align}\label{eqo2}
\left. \frac{d P_r}{d t}\right |_{\rm odd} = -16\pi \frac{\partial H_{\rm int}^{\rm odd}}{\partial R} 
=& 16 \pi \left [ h_{0,r}\Xi +\frac{2 h_{1,r}P^r\Xi+h_{2,r}\Pi}{2\,P^0}\right ]  \nonumber \\
+&16 \pi \left [ 2 h_1P^r\Xi+h_2\Pi\right ]\left [ \frac{1}{P^0}\right ]_{,r} \\
\label{eqo3}
\left.\frac{d P_{\phi}}{d t}\right|_{\rm odd} = -16\pi \frac{\partial H^{\rm odd}_{\rm int}}{\partial \Phi}  
=& 16 \pi  \,h_0P_{\phi}X^{\phi}_{,\phi}\nonumber\\
+&16\pi \left [ \frac{2 h_1P^rP_{\phi}X^{\phi}_{,\phi}+h_2W^{\phi\phi}_{,\phi}(P_{\phi})^2}{2 P^0}\right ] 
\end{align}
\end{subequations}
Note that such a term exists for each $(l,m)$ with $l\ge 2$.

From the above equations of motions it is clear that the effect of odd-parity perturbations the test particle's motion is determined once we know $h_0,h_1,h_2$ or $h_0,k_1,k_2$ and their spatial derivatives at $r=R(t)$; here $h_0$ and $k_2$ are related with the actual gauge choice and $k_1$ is gauge invariant. 
If we track back to the wave equation for $\psi_{\rm RW}$ or $k_1$, it is easy to see that $k_1$ must be discontinuous at $r=R(t)$ in order to obtain a source function $\delta'(r-R(t))$. On the other hand, the equation of motion for $\left. d P_r/d t\right |_{\rm odd}$ contains a term proportional to $k_{1,r}$. That means that this equation of motion is singular because it contains $\delta(r-R(t))$. 
This means  the full metric perturbation is singular at the point particle's location, and directly applying full metric perturbation to the particle's equations of motion will introduce divergence. 
One has to apply a regularization scheme before one can use these equations for computation. 
This scheme must regularize gauge-invariant quantity $k_1$ as well as gauge-dependent terms $h_2, h_0$, since they all enter the particle's equation of motion. We will discuss possible regularization methods  in Sec. \ref{sec4}. 

\subsection{Even Parity $(l \ge 2)$}
Even-Parity metric perturbations are described in terms of the scalar harmonics $Y^{lm}$, vector harmonics $Z^{lm}_A$ and tensor harmonics $U^{lm}_{AB}$ and $V^{lm}_{AB}$ [see Eqs.~\eqref{eqg1} and \eqref{eqng2}. For later convenience, we  define the following quantities,
\begin{eqnarray}
\Pi_1=U^{lm}_{AB}P^AP^B, \, \Pi_2=V^{lm}_{AB}P^AP^B, \,  \Xi'=Z^{lm}_A P^A 
\end{eqnarray}
which are contractions of the harmonics with angular components of the momentum.   Even-parity metric perturbation fields include: 
\begin{subequations}
\begin{align}
N'=&-\frac{1}{2}f^{1/2} H_0 Y^{lm}, \; N'_r=H_1 Y^{lm}, \; N'_A = h^*_0 Z^{lm}_A \;  \\
h^{lm}_{rr} =& \frac{ H_2 }{f}Y^{lm},\; h^{lm}_{rA} = h^*_1 Z^{lm}_A, \; h^{lm}_{AB} = r^2( K U^{lm}_{AB}+G V^{lm}_{AB})  
\end{align}
\end{subequations}
The $(l,m)$ component of the even-parity Hamiltonian is 
\begin{align}
\label{Hinteven}
 H_{\rm int}^{\rm even} =& \int dr 
 \int d^2\Omega \ \delta(r-R(t))\delta^{(2)}(\Omega-\tilde\Omega(t)) \nonumber\\
&\qquad \quad \bigg[ -h_0^*\Xi'-H_1Y^{lm}P^r + H_0 Y^{lm}P_0/2 \nonumber \\
& \qquad\quad -  \frac{f^{-1} (P^r)^2 H_2 Y^{lm}+2\,h_1^*P^r\Xi'}{2 P^0} \nonumber\\
& \qquad \quad -\frac{+r^2(K\Pi_1+G\Pi_2)}{2P^0}\bigg] 
\end{align}
Here again we have used $\Omega$ to represent $(\theta,\phi)$, and $\tilde\Omega(t)$ to represent $(\Theta(t),\Phi(t))$, and have defined $P_0=-f P^0$. In addition,  $h^*_0$, $H_1$ and $H_0$ are the lapse and shift perturbations, they serve as Lagrange multipliers in the Hamiltonian; $K$, $H_2$, $h^*_1$ and $G$ are $3$-metric perturbations, they couple with the point-particle dynamical variables at its location, sourcing the interaction between the field and the test mass. The relation between $K$, $H_2$, $h^*_1$, $G$ and Moncrief's $q_1$, $q_2$, $q_3$, $q_4$ are shown in Appendix \ref{ap2}.  We recall the subtlety involving $\delta $ function and its derivative mentioned below Eq.~\eqref{Hintodd}, and note that all terms in the integrand of Eq.~\eqref{Hinteven}, with the exception of the $\delta$ function, only depend explicitly on $(r,\theta,\phi)$, but not on $(R,\Theta,\Phi)$. 

By varying $h^*_0, H_1, H_0$, we can obtain the new constraint equations:
\begin{subequations}\label{eqconstrainte}
\begin{align}
q_2 =& -16\pi P_0 Y^{lm} \delta(r-R(t))  \\
\pi_4 =& 16\pi Y^{lm} P^r \delta(r-R(t))  \\
\pi_3 =& 8\pi r^2\Xi'\delta(r-R(t))+8\pi Y^{lm}[r^2P^r\delta(r-R(t))]_{,r} 
\end{align}
\end{subequations}
This means $q_2$, $\pi_3$ and $\pi_4$ are all  divergent at the test particle's location and vanish everywhere else.  

From $H_{\rm int}^{\rm even}$~[Eq.~\eqref{Hinteven}], the evolution equation for $\pi_1$ gains the additional term of 
\begin{align}
\left. \frac{\partial \pi_1}{\partial t}\right |_{\rm add}&=-\frac{\lambda+1}{r f}\left. \frac{\partial \pi_2}{\partial t}\right |_{\rm add}+\left. \frac{\partial \pi_2}{\partial t}\right |_{\rm add,r} \nonumber \\
&+\frac{4\pi(P^r)^2Y_{lm}}{P^0f^2}\delta(r-R(t))
\end{align}
and evolution equation of $\pi_2$ gains 
\begin{align}
\left. \frac{\partial \pi_2}{\partial t}\right |_{\rm add}& =\frac{4\pi[(P^r)^2Y_{lm}\Lambda+2(P^0)^2r^2f^2\Pi]}{P^0f(1+\lambda)\Lambda}\delta(r-R(t)) \nonumber \\
&+\frac{16\pi f}{(\lambda+1)\Lambda}\left [ \frac{r(P^r)^2Y_{lm}}{2fP^0}\delta(r-R(t))\right ]'
\end{align}
Similar to the odd-parity case, the evolution equations for $\pi_3,\,\pi_4$ (up to linear order) are consistent with the constraint equations Eq.~(\ref{eqconstrainte}).

 Combing the evolution equation for $q_1,q_2,\pi_1,\pi_2$ and the constraint equations, we will find that the gauge-invariant field $\psi_{\rm ZM}$ satisfies a wave equation with source term coming from the point particle:
\begin{equation}\label{eqso}
\left [ -\frac{\partial^2}{\partial t^2}+ \frac{\partial^2}{\partial r^{*2}}-V^l_{\rm even}(r) \right ] \psi_{\rm ZM}(r,t) = S_{\rm even}(r,t)
\end{equation}
with the potential $V^l_{\rm even}$  given by
\begin{align}
V^l_{\rm even} =\frac{4 f}{r^2 \Lambda^2}&\bigg [ 2\lambda^2\left(\lambda+1+\frac{3 M}{r}\right) \nonumber\\
&+\frac{18M^2}{r^2}\left(\lambda+\frac{M}{r}\right)\bigg]
\end{align}
 and the source term $S_{e}$ give by 
 \begin{align}
 \label{source}
 S_{\rm e} &= \frac{2}{(\lambda+1)\Lambda}\left \{ r^2 f (f^2 \frac{\partial}{\partial r}Q^{tt}-\frac{\partial}{\partial r}Q^{rr})+r(\Lambda/2-f)Q^{rr} \right.\nonumber \\
 &-   \left. \frac{2 f^2}{r\Lambda}[\lambda(\lambda-1)r^2+(4\lambda-9)Mr+15M^2]Q^{tt} \right \} \nonumber \\
&+ \frac{ 2r f^2}{(\lambda+1)\Lambda} Q^{\flat}+\frac{4f}{\Lambda}Q^r-\frac{f}{r}Q^{\sharp}
   \end{align}
Here the $Q$'s are master functions describing spherical-harmonic decompositions of point mass stress energy tensor. They are defined by
\begin{subequations}
\label{eq:Qs}
\begin{align}
Q^{tt} & = 8\pi \int T^{tt} Y^{lm*}d\Omega = \frac{8\pi  P^0}{r^2}\delta[r-R(t)]Y^{lm*}[\Omega(t)],  \\
Q^{rr} & = 8\pi \int T^{rr} Y^{lm*}d\Omega =  \frac{8\pi(P^r)^2\delta[r-R(t)]Y^{lm*}[\Omega(t)]}{r^2 P^0},  \\
Q^r &= \frac{8 \pi r^2}{\lambda+1} \int T^{rA}Z^{lm*}_A d\Omega =  \frac{8 \pi P^r \Xi'}{(\lambda+1)P^0} \delta[r-R(t)],  \\
Q^{\flat} &= 8\pi r^2 \int T^{AB}U^{lm*}_{AB}d\Omega =\frac{8 \pi\Pi_1}{P^0}\delta[r-R(t)],  \\
Q^{\sharp} &= \frac{8 \pi r^4}{\lambda(\lambda+1)}\int T^{AB} V^{lm*}_{AB}d\Omega = \frac{8 \pi r^2 \Pi_2\delta[r-R(t)]}{\lambda(\lambda+1)P^0}.
\end{align}
\end{subequations}
The source term in Eqs.~\eqref{source} and~\eqref{eq:Qs} agrees with the previous derivation of Martel and Poisson~\cite{Karl,Karl2}, and here we have adopted their notation.

In addition to the source term in the constraint equations and the field evolution equations,  the particle-field interaction Hamiltonian also generate additional terms in the particle's equation of motion, which causes radiation reaction.   These terms can be obtained by varying the interaction Hamiltonian with respect to point mass dynamical variables, in a similar manner as the odd parity case, for the canonical coordinate
\begin{widetext}
\begin{subequations}
\begin{align}
\left. \frac{d R}{d t}\right |_{\rm even}  = 16\pi \frac{\partial H_{\rm int}}{\partial P_r} &=  -16 \pi \left [H_1 Y^{lm} f+ \frac{  H_0 Y^{lm} P^r}{2 P^0}+\frac{f P_rH_2Y^{lm}}{P^0}+\frac{f h^*_1\Xi'}{P^0}\right] \nonumber\\
&+ 16\pi P_r\frac{(P^r)^2/f H_2Y^{lm}+2h^*_1 \Xi'P^r+r^2(K \Pi_1+G \Pi_2)}{2(P^0)^3} \\
\left. \frac{d \Phi}{d t}\right |_{\rm even} = 16\pi \frac{\partial H_{\rm int}}{\partial P_{\phi}} 
 &= 16 \pi \left \{-h^*_0Z^{\phi}_{lm} -\frac{ H_0Y^{lm}P_{\phi}}{2r^2P^0\sin^2{\theta}} +\frac{P_{\phi}}{2 f r^2\sin^2{\theta}(P^0)^3} \left [ f^{-1} (P^r)^2 H_2 Y^{lm}+2\,h_1^*P^r\Xi'+r^2(K\Pi_1+G\Pi_2)\right ] \right \} \nonumber \\
&  -16\pi\frac{h^*_1P^rZ^{\phi}+r^2(K U^{\phi\phi}P_{\phi}+G V^{\phi\phi}P_{\phi})}{P^0}
\end{align}
and their conjugate momentum
\begin{align}
\left. \frac{d P_r}{d t}\right |_{\rm even} = -16\pi \frac{\partial H_{\rm int}}{\partial R} 
=& 16 \pi \left \{ \Xi'\frac{\partial h^*_0(R)}{\partial R}+Y^{lm}P_r\frac{\partial  (f H_1(R))}{\partial R}-\frac{1}{2}Y^{lm}\frac{\partial ( H_0(R)P_0)}{\partial R}\right \} \nonumber \\
&+16\pi \left \{ \frac{1}{2P^0}\left [ (P_r)^2Y^{lm}\frac{\partial (f H_2(R))}{\partial R}+2P_r \Xi' \frac{\partial (f h^*_1(R))}{\partial R}+\Pi_1 \frac{\partial (R^2 K(R))}{\partial R}
+\Pi_2 \frac{\partial (R^2 G(R))}{\partial R}\right ]\right \} \nonumber \\
&- \frac{8\pi}{(P^0)^2} \frac{\partial P^0}{\partial R} \left[ f^{-1} (P^r)^2 H_2 Y^{lm}+2\,h_1^*P^r\Xi'+r^2(K\Pi_1+G\Pi_2)\right ]  \\
\left. \frac{d P_{\phi}}{d t}\right |_{\rm even} = -16\pi \frac{\partial H_{\rm int}}{\partial \phi} 
=&  16\pi \left[ h^*_0 \frac{\partial \Xi'}{\partial \phi}+H_1P^r \frac{\partial Y^{lm}}{\partial \phi}-\frac{1}{2}H_0P_0\frac{\partial Y^{lm}}{\partial \phi} \right ]\nonumber \\
&+\frac{8\pi}{P^0}\left [ f^{-1}(P^r)^2H_2\frac{\partial Y^{lm}}{\partial \phi}+2h^*_1P^r\frac{\partial \Xi'}{\partial \phi}+r^2G\frac{\partial \Pi_2}{\partial \phi}+r^2K\frac{\partial \Pi_1}{\partial \phi}\right ]
\end{align} 
\end{subequations}
\end{widetext}
We have defined 
 \begin{align}
 P^0(R) =\sqrt{(P_r)^2+\frac{(P_{\phi})^2}{R^2(1-2M/R)\sin^2{\theta}}+\frac{m^2}{1-2M/R}}
 \end{align}
This set of equations, together with the even-parity wave equation (\ref{eqso}) and the odd-parity equations \eqref{eqol}, \eqref{eqo1}, \eqref{eqo2} and \eqref{eqo3},  form a complete set of self-consistent evolution equations for both the point particle and the metric-perturbation fields. 

Similar to the odd parity case, the even parity equations of motion also has a divergence problem. Because the wave equation~\eqref{eqso} for  $\psi_{\rm ZM}$ contains a source term as singular as $\delta'(r-R(t))$, $\psi_{\rm ZM}$, or $q_1$, must be discontinuous at the point particle's location. According to the relation between  $H_2, G, h^*_1, K$ and $q_1,q_2,q_3,q_4$ shown in Appendix \ref{ap1},  $K$ contains a $\delta(r-R(t))$-type term and $H_2$ even contains a $\delta'(r-R(t))$-type term. This means terms added to the particle's equation of motion are all singular at the particle's location.  As a result, one has to {\it regularize} these equations of motion before they can be used for actual computations. 

 \subsection{Monopole and dipole perturbations}

Even though there are no gauge-invariant perturbations for these low-$l$ components, metric perturbations at these orders do couple to the particle.  The particle's perturbation to fields at these orders have been solved explicitly by Detweiler and Poisson~\cite{poisson2}, while their back-action to the particle's canonical equations can be obtained from expressions obtained for $l\ge 2$, simply removing those terms that do not exist in these low $l$'s. 
 
  \section{Regularization of  Test Particle Equation of Motion}\label{sec4}

In order to obtain regular equations of motion for the point particle, we must carry out a {\it regularization procedure} that appropriately removes the divergences from the metric perturbation fields. While we have not been able to find a stand-alone regularization procedure in the 3+1 picture, currently existing regularization schemes can be adapted to our formalism.   In this section, we shall outline, but not carry out, the procedure with which such a regularization could be done.  

\subsection{General Discussion}

In particular, we shall discuss how the Detweiler-Whiting (DW) singular-regular decomposition~\cite{detweiler, detweiler2} approach can be used to regularize our canonical equations of motion. 
In the DW approach, metric perturbation field in a small but finite region around the point particle is decomposed into the sum of a regular piece (superscript ``R'') and a singular  piece (superscript ``S''):
\begin{equation}
h_{\mu\nu} = h^R_{\mu\nu}+h^S_{\mu\nu}\,.
\end{equation}
The singular piece $h^S_{\mu\nu}$ corresponds to the deformed Schwarzschild solution around the small test mass as seen by a locally free falling observer on the background spacetime --- it is singular as we approach the location of the point particle; the regular piece $h_R$ satisfies the linearized vacuum Einstein's Equation  and is everywhere regular (although it  {\it does not} satisfy the out-going boundary condition at the null infinity and the down-going boundary condition at the future horizon).  It is shown that $h^S_{\mu\nu}$ is the appropriate singularity to remove, and the point particle should travel along a geodesic of the perturbed spacetime that differs from the background by $h^R_{\mu\nu}$. 
%

DW has shown that $h^S_{\mu\nu}$ can be approximated analytically in a local normal coordinate system built around the particle --- such as the one  introduced by Thorne and Hartle \cite{THZ} and  developed to higher orders by Zhang \cite{zhang} (usually referred to as the THZ coordinate system).  
Another approach towards obtaining $h_{\mu\nu}^S$ is through the Hadamard singular Green function, as carried out by Hass and Poisson~\cite{roland} as well as Warburton et al.~\cite{warburton}.   The computation for $h^S_{\mu\nu}$  is carried out as an expansion in the proper distance away from the particle --- and depending on the order to which this expansion is carried out, the corresponding $h_{\mu\nu}^R$ will only have a finite order of smoothness. 

Among components of $h_{\mu\nu}^S$, $h_{tt}^S$ is a lapse perturbation, $h_{tr}^S$ and $h_{tA}^S$ are shift perturbations, while $h_{rr}^S$, $h_{rA}^S$, and $h_{AB}^S$ are 3-metric perturbations.  One can carry out $(l,m)$ decompositions of these quantities, using the appropriate harmonics, to obtain the singular pieces of our odd-parity metric-perturbation fields $(h_0^S,h_1^S,h_2^S)$ and even-parity metric-perturbation fields $(H_0^S, H_1^S,h_0^{*S},h_1^{*S},H_2^S,K^S,G^S)$.  In this way, the singular metric-perturbation fields come with a choice of gauge (through the singular pieces for lapse and shift) as well as 3-metric perturbations, around the worldline of the point particle.  The $(l,m)$-decomposition coefficients of the singular metric fields are also referred to as {\it regularization parameters}.  

It is anticipated that the mode-decomposed versions of these singular metric-perturbations fields should in general be discontinuous or singular at the radial location of the particle --- but it is exactly these singularities that will cancel with the ones we obtain for the full perturbations (i.e., $h_{\mu\nu}^{\rm full}$), yielding 
\begin{equation}
\label{subtraction}
h_{\mu\nu}^R = h_{\mu\nu}^{\rm full} - h_{\mu\nu}^S
\end{equation}
which are regular.

More specifically in the 3+1 approach, we must first obtain the full metric (including lapse, shift and 3-metric perturbations), and then subtract the singular piece --- resulting in the regular piece.  A subtlety here is the choice of gauge: we obtain $h_{\mu\nu}^{\rm full}$ using a particular choice of lapse and shift perturbations, and the arbitrariness of the choice suggests that the subtraction~\eqref{subtraction} will yield a regular result only if the full metric and the singular metric are computed in gauges that are related to each other through a smooth transformation in the region near the particle. 


Note that the singular field is only defined in a region around the point particle --- because the normal coordinate system (e.g., the THZ coordinate system), as well as the Hadamard decomposition of the Green function, is only valid within a distance away from the particle that is comparable to space-time curvature.   This does not prevent us from obtaining a regularized set of equations of motion for the point particle, because for that we will only need to obtain $h_{\mu\nu}^R$ around the location of the particle.

However, this has lead Vega and Detweiler (VD)~\cite{Ian1} to develop a slight variant of the DW regularization approach, which further simplifies the regularization procedure.  VD first assumed that we can obtain an $h_{\mu\nu}^S$ that has a definition everywhere in the spacetime, although this definition is physically meaningful only around the particle.  They then proposed the application of a window function $W$, which is very flat around the location of the particle, but decays rapidly towards the horizon and infinity.  In this way, if one defines an effective regular field, or {\it effective field} for short, 
\begin{equation}
\tilde h_{\mu\nu}^R \equiv h^{\rm full}_{\mu\nu} - W h^S_{\mu\nu}
\end{equation}
then the effective field $\tilde h_{\mu\nu}^R$ satisfies a wave equation with a regular source (the full source subtracted by the result obtained by inserting $\tilde{h}^S_{\mu\nu}\equiv Wh^S_{\mu\nu}$ into the wave equation), as well as the out-going boundary condition at the future null infinity and the down-going boundary condition at the future horizon. 

In this paper, we shall discuss how the effective-source approach can be adapted to our 3+1 Hamiltonian formalism --- in even- and odd-parity cases. 


\subsection{Odd parity}

Odd parity effective fields are $\tilde{h}_{0R}, \tilde{h}_{1R}, \tilde{h}_{2R}$, and their smoothness depends on the quality of our approximations for $h_{0S}$, $h_{1S}$ and $h_{2S}$.   We shall refer the regular field to be $n$-th order smooth if it has a smooth $n$-th order derivative.  Right now, singular field is available for the regular field to have $4$-th order smoothness \cite{anna}. Let us first assume the order of smoothness is not an issue (e.g., assuming the singular piece to be available up to a rather high order), and later discuss options when the order of smoothness is limited.

\subsubsection{An algebraic gauge}

%
 %
 Out of the three metric quantities, one can construct a gauge invariant quantity --- the Regge-Wheeler (RW) function, and the rest two degrees of freedom are fixed by one gauge choice and one constraint equation.  First consider the gauge invariant quantity, its effective regularized piece $\tilde{\psi}_{\rm RWR}$  is  given by 
\begin{equation}
\label{eqrwr}
\tilde{\psi}_{\rm RWR} = \frac{f}{r} \left [ \tilde{h}_{1R}+\frac{1}{2}\left ( \frac{\partial \tilde{h}_{2R}}{\partial r}-\frac{2}{r}\tilde{h}_{2R}\right )\right ]
\end{equation}
while its effective singular piece $\tilde{\psi}_{\rm RW}^S(r,t)$  is given similarly by $\tilde h_{1S}$, $\tilde h_{2S}$  (which are singular field components multiplied by the window functions). 

The effective RW function satisfies the same wave equation as before [See Eq.~\eqref{eqol}] but with a new source
\begin{equation}\label{eqefo}
\left [ -\frac{\partial^2}{\partial t^2}+ \frac{\partial^2}{\partial r^{*2}}-V_{\rm odd}(r) \right ] \tilde{\psi}_{\rm RWR}(r,t) = S_{\rm odd R}(r,t)
\end{equation}
where the new source $S_{\rm odd}^{R}$ is simply the effective source, given by
\begin{equation}
S_{\rm odd R}(r,t) = S_{\rm odd}(r,t)-\left [ -\frac{\partial^2}{\partial t^2}+ \frac{\partial^2}{\partial r^{*2}}-V_{\rm odd}(r) \right ] {\tilde\psi}_{\rm RWS}(r,t) 
\end{equation}
Given enough smoothness on the singular piece, a smooth enough $\tilde\psi_{\rm RWR}$ can be obtained by solving Eq.~\eqref{eqefo} and imposing the out-going and down-going boundary conditions and infinity an the horizon.  This $\tilde \psi_{\rm RWR}$ can then be used to construct the rest of the gauge-dependent fields, imposing the gauge condition of, for example, $\tilde{h}_{1R}=0$.  We then obtain regular values for all the metric perturbation fields, as well as their derivatives, at the location of the point particle and will be able to drive its motion.  To carry out this computation, we will need the regular field to be 2-order smooth.  

To be more specific, we can  always do the coordinate transformation similar to Eq.~(\ref{eqocoord}) to shift $\tilde{h}_{1R}$ to $0$. The gauge transformation function is given by
\begin{subequations}\label{eqregtran}
\begin{align}
& C^{lm}_{,r}-\frac{2}{r}C^{lm}=-\tilde{h}_{1R}\rightarrow \,C^{lm}=-r^2\int dr \frac{\tilde{h}_{1R}}{r^2}  \\
& x'_A =x_A+\sum_{lm} C^{lm}X^{lm}_A=x_A-r^2\int dr \frac{\tilde{h}^{\rm odd}_{rA}}{r^2}
\end{align}
\end{subequations}

After the gauge transformation, according to Eq.~(\ref{eqocoord}), the new $\tilde{h}'_{0R}, \tilde{h}'_{1R}, \tilde{h}'_{2R}$ are
\begin{subequations}
\label{eqregh}
\begin{align}
\tilde{h}'_{1R}= &0, \\ 
\tilde{h}'_{2R}=&\tilde{h}_{2R}+2r^2\int dr \frac{\tilde{h}_{1R}}{r^2}, \\
 \tilde{h}'_{0R}=&\tilde{h}_{0R}-2r^2\int dr \frac{\partial_t\tilde{h}_{1R}}{r^2}
\end{align} 
\end{subequations}

Suppose the original effective field $\tilde{h}_R$ is a $C^n$ function on the test mass's worldline, the coordinate transformation must be $C^n$ smooth (Eq.~(\ref{eqregtran})) and the new effective field is $C^{n-1}$ smooth (Eq.~(\ref{eqregh})). Therefore $\tilde{h}_{1R}=0$ is also a viable gauge for evolution because it can be smoothly transformed from Lorenz gauge. For $n\ge2$ the spatial derivative of the metric components would still be continuous. By imposing the $\tilde{h}_{1R}=0$ algebraic gauge condition, $\tilde{h}_{2R}$ can be immediately obtained by solving 
\begin{equation}
\frac{\partial \tilde{h}_{2R}}{\partial r}-\frac{2}{r}\tilde{h}_{2R}=\frac{2r}{f}\tilde{\psi}_{\rm RWR}
\end{equation}

As $\tau_2$ is fixed by the constraint equation Eq.~(\ref{eqocon}) and $\tau_1$ can be obtained by solving Eq.~(\ref{eqsovtau1}),  it is then straight forward to obtain $\tilde{h}_{0R}$ through Eq.~(\ref{eqmo1})
\begin{equation}
\tilde{h}_{0R}=-\tilde{h}_{0S}-\frac{1}{2}\frac{\partial (\tilde{h}_{2R}+\tilde{h}_{2S})}{\partial t}+\frac{f r^2}{2\lambda(\lambda+1)}\left [\tau_2-\frac{1}{2r^2}\frac{\partial(r^2\tau_1)}{\partial r} \right]
\end{equation}

Compared to the Lorenz gauge condition (in the following section), computing metric perturbations in this algebraic gauge is relatively easier although the effective fields are 1 order worse in smoothness.


\subsubsection{Fixing Lorenz Gauge}

Another way to ensure the smoothness of the regular field is to resort to the known conclusion that if we keep the full field in the Lorenz gauge, the existing  $n=1$ singular field should be sufficient.  This has been demonstrated by Refs.~\cite{mino, barack, detweiler2}.  

The Lorenz gauge condition,
\begin{equation}\label{eqoharreq}
\triangledown^{\mu} \bar{h}_{\mu\nu} =0
\end{equation}
where $\bar{h}_{\mu\nu}$ is the trace reversed metric perturbation $\bar{h}_{\mu\nu}=h_{\mu\nu}-1/2 \,g_{\mu\nu}h_{\alpha\beta}g^{\alpha\beta}$, converts into 
\begin{equation}
r \frac{\partial h_0}{\partial t}+2 f \left (\frac{M}{r}-1 \right )h_1-f^2 r \frac{\partial h_1}{\partial r}+\lambda f h_2=0
\end{equation}
for the $(l,m)$ odd-parity perturbation fields. As we break this into singular and regular pieces, we obtain
\begin{equation}\label{eqharo}
r \frac{\partial \tilde{h}_{0R}}{\partial t}+2 f \left (\frac{M}{r}-1 \right )\tilde{h}_{1R}-f^2 r \frac{\partial \tilde{h}_{1R}}{\partial r}+\lambda f \tilde{h}_{2R}=A
\end{equation}
for the effective regular field components, where $A$ is given by 
\begin{equation}
A=-r \frac{\partial \tilde{h}_{0S}}{\partial t}-2 f \left (\frac{M}{r}-1 \right )\tilde{h}_{1S}+f^2 r \frac{\partial \tilde{h}_{1S}}{\partial r}-\lambda f \tilde{h}_{2S}
\end{equation}
Combining Eqs.~(\ref{eqrwr}), (\ref{eqharo}) and (\ref{eqmo1}), we have a set of first-order differential equations for $\tilde{h}_{0R}$ and $\tilde{h}_{2R}$
\begin{align}\label{eqharoevo}
\partial_t\left[\begin{array}{c}
\tilde{h}_{0R} \\
\tilde{h}_{2R} \\
\end{array}
\right]
= &\left[
\begin{array}{c c}
M_{11} & M_{12} \\
 M_{21}& M_{21} \\
\end{array}
\right]
\left[\begin{array}{c}
\tilde{h}_{0R} \\
\tilde{h}_{2R} \\
\end{array}
\right]
+\left[\begin{array}{c}
N_1 \\
N_2 \\
\end{array}
\right] 
\end{align}
with 
\begin{subequations}
\begin{align}
 M_{11} = & M_{22}= 0 \\
 M_{12} = & -\frac{\lambda f}{r} +\left[ f^2 \frac{\partial}{\partial r}-\frac{2 f}{r}\left (\frac{M}{r}-1\right )\right] \left[ \frac{1}{r}-\frac{\partial}{2\partial r}\right]  \\
M_{21}= & -2,
\end{align}
\end{subequations}
and
\begin{subequations}
\begin{align}
 N_1= & \frac{A}{r}+\left[ f^2 \frac{\partial}{\partial r}-\frac{2 f}{r}\left (\frac{M}{r}-1\right )\right]\tilde{k}_{1R}, \\
  N_2 =&  \frac{f r^2}{\lambda(\lambda+1)}\left (\tau_2-\frac{1}{2r^2}\frac{\partial(r^2 \tau_1)}{\partial r}\right )
\end{align}
\end{subequations}
Here $\tau_2$ is fixed by the constraint equation Eq.~(\ref{eqocon}) and $\tau_1$ can be obtained by solving Eq.~(\ref{eqmo1})
\begin{align}\label{eqsovtau1}
&\frac{\partial (\tilde{k}_{1R}+\tilde{k}_{1S})}{\partial t}\nonumber\\
 =&\frac{\tau_1}{2(\lambda+1)}+\frac{r^2}{2\lambda(\lambda+1)}\frac{\partial}{\partial r}\left [ f (\tau_2-\frac{1}{2r^2}\frac{\partial (r^2 \tau_1)}{\partial r})\right] 
\end{align}

At the initial time slice $t=0$, we can impose the initial gauge condition that $\tilde{h}_{0R}(t=0) = \tilde{h}_{2R}(t=0)=0$ and Eq.~(\ref{eqharoevo}) determines the gauge condition evolution later on. Given $\tilde{h}_{0R}, \tilde{h}_{2R}$, Eq.~(\ref{eqrwr}) determines the value for $\tilde{h}_{1R}$ and therefore we can obtain the full set of regularized odd-parity field.

\subsection{Even parity}

For even parity, we will follow similar procedures as the odd-parity case. Here we are dealing with seven effective field quantities: $\tilde{K}_R, \tilde{G}_R,\tilde{H}_{2R},\tilde{H}_{1R},\tilde{H}_{0R},\tilde{h}^*_{1R}, \tilde{h}^*_{0R}$ compared to three field quantities in the odd-parity case. Out of these seven quantities, one can construct one gauge invariant perturbation quantity -- Zerilli-Moncrief quantity and the rest six degrees of freedom are fixed by three gauge conditions and three constraint equations. The regular piece of the Zerilli-Moncrief function  is given by
\begin{eqnarray}
\tilde{\psi}_{ZMR} &=& \frac{r}{\lambda+1}\left[ \tilde{K}_R+\frac{2 f}{\Lambda}\left (\tilde{H}_{2R}-r\frac{\partial \tilde{K}_R}{\partial r}\right )\right ]  \nonumber \\
&+& \frac{2 f }{\Lambda}\left (\frac{r^2 \partial \tilde{G}_R}{\partial r}-2 \tilde{h}_{1R} \right )
\end{eqnarray}
 It satisfies the following wave equation 
\begin{equation}
\left [ -\frac{\partial^2}{\partial t^2}+ \frac{\partial^2}{\partial r^{*2}}-V_{\rm even}(r) \right ] \tilde{\psi}_{ZMR}(r,t) = S_{eR}(r,t)
\end{equation}
with the effective source term $S_{eR}$ given by
\begin{equation}
S_{eR}(r,t) = S_{e}(r,t)-\left [ -\frac{\partial^2}{\partial t^2}+ \frac{\partial^2}{\partial r^{*2}}-V_{\rm even}(r) \right ] \tilde{\psi}_{ZMS}(r,t) 
\end{equation}
With outgoing wave boundary condition at spatial infinity and black hole horizon, one can solve the wave equation and obtain the numerical value for $\tilde{\psi}_{ZMR}$ or $\tilde{q}_{1R}$. On the other hand, the effective field $\tilde{q}_{2R}$ is fixed by the constraint equation
\begin{eqnarray}
\tilde{q}_{2R} = q_2-\tilde{q}_{2S} = -16\pi P_0Y^{lm}\delta(r-R(t))-\tilde{q}_{2S}
\end{eqnarray}

Similar to the odd parity case, if quality of the singular field is high enough, we can simply set the additional lapse $\tilde H_{0R}$ and shifts $(\tilde H_{1R},\tilde h^*_{0R})$ to zero, or to use an algebraic gauge for the gauge-dependent fields.  Also similar to the odd parity, one way to limit our requirement for smoothness is to apply Lorenz gauge condition Eq.~(\ref{eqoharreq}),  similar to what we did for the odd-parity case. In this case these Lorenz gauge conditions are given by
\begin{subequations}
\begin{align}
0 &= (\lambda+1)\frac{h^*_0}{r^2}+\left( \frac{M}{r}-1\right)\frac{H_1}{r}-\frac{f}{2}\frac{\partial H_1}{\partial r} \nonumber\\
&+\frac{1}{4}\frac{\partial (H_0+H_2+2K)}{\partial t} \\
0&= \frac{M}{r}H_0-2(\lambda+1)\frac{f h^*_1}{r}+\left( 2-\frac{3M}{r}\right)H_2-2fK \nonumber\\
&+\frac{1}{2}rf\frac{\partial (H_0+H_2-2K)}{\partial r}-r\frac{\partial H_1}{\partial t} \\
0&=f \left [ \left ( \frac{M}{r}-1\right)h^*_1+\left( \frac{\lambda}{2}G+\frac{H_2-H_0}{4}\right)r\right]+\frac{r f^2}{2}\frac{\partial h^*_1}{\partial r} \nonumber\\&-\frac{r}{2}\frac{\partial h^*_0}{\partial t}
\end{align}
\end{subequations}
 Combing the above equations with Appendix \ref{ap1}, Eq.~(\ref{eqe}) as well as the constraint Eq.(\ref{eqconstrainte}), one can write down the evolution equation for $\tilde{G}_R, \tilde{h}^*_{1R}, \pi_2,\tilde{H}_{1R},\tilde{h}^*_{0R}$ and a combination of effective field functions $I_R=\tilde{H}_{0R}+\tilde{H}_{2R}+2\tilde{K}_R$. We also correspondingly define $\,I_S=\tilde{H}_{0S}+\tilde{H}_{2S}+2\tilde{K}_S$.
\begin{equation}\label{eqhareevo}
\partial_t\left[\begin{array}{c}
\tilde{G}_{R} \\
\tilde{h}^*_{1R} \\
\pi_2 \\
I_R \\
\tilde{H}_{1R} \\
\tilde{h}^*_{0R} \\
\end{array}
\right]
={\mathbf{ M'}}\left[\begin{array}{c}
\tilde{G}_{R} \\
\tilde{h}^*_{1R} \\
\pi_2 \\
I_R \\
\tilde{H}_{1R} \\
\tilde{h}^*_{0R} \\
\end{array}
\right]+\mathbf{N'}
\end{equation}
 Non-zero components of the matrix $\mathbf{M'}$ in Eq.~\eqref{eqhareevo} are given by
\begin{subequations}
\begin{align}
& M'_{13}=\frac{2 f}{r^2}, \; M'_{16} =\frac{2}{r^2};  \\
& M'_{23}=2f\partial_r-\frac{2}{r}\left(f-\frac{M}{r}\right),\; M'_{25}=1, \; M'_{26}=\partial_r-\frac{2}{r}; \\
& M'_{31}=-\frac{r^2}{2} f \partial^2_r,\quad M'_{32}=\frac{f}{r}(2-r\partial_r),\quad M'_{34}=-\frac{1}{2}; \\
&M'_{45}=-4\left(\frac{M}{r^2}-\frac{1}{r}\right)+2 f \partial_r,\quad M'_{46}=-\frac{4(\lambda+1)}{r^2} r; \\
&M'_{51}=\frac{M}{r^2}+\frac{f }{2}\partial_r ,\quad M'_{52}=-\frac{4 M f}{r}\partial_r ,\quad \nonumber\\& M'_{54}=-\frac{2f}{r^3}(4M+r+\lambda r) ; \\
&M'_{61}= \frac{\lambda f}{r}-(3-5M/r)f \partial_r -r f^2\partial_r^2,\quad M'_{62}=3 f^2\partial_r,\nonumber\\
&M'_{64}=-\frac{f}{2r}.
\end{align}
\end{subequations}
Those of $\mathbf{N'}$  given by 
\begin{subequations}
\begin{align}
N'_1 &= \frac{f \pi_3}{r^2 \lambda(\lambda+1)}+\frac{f \Lambda \pi_1}{r \lambda}+\frac{2\tilde{h}^*_{0S}}{r^2}-\frac{\partial \tilde{G}_S}{\partial t},\\
 N'_2& =\frac{\pi_4}{2(\lambda+1)}-2f\pi_1+\tilde{H}_{1S}+r^2\left(\frac{\tilde{h}^*_{0S}}{r^2}\right)_{,r}-\frac{\partial \tilde{h}^*_{1S}}{\partial t} \\
N'_3&=\frac{\lambda}{r \Lambda^2}q_1-\frac{\lambda f}{(\lambda+1)\Lambda^2}(q_2-q_{1,r})-\frac{M}{2 r}(r \frac{\partial \tilde{G}_S}{\partial r}-\frac{2}{r}\tilde{h}^*_{1S})\nonumber\\
&+ \frac{M(2 q_2-q_{1,r})}{2(\lambda+1)\Lambda r}-\frac{\tilde{H}_{0S}}{2} +\frac{4\pi}{P^0}\frac{(P^r)^2Y^{lm}}{f(\lambda+1)}\delta(r-R(t)) \nonumber\\
&+\frac{8\pi f}{\Lambda(\lambda+1)}\left [ \frac{(P^r)^2Y^{lm}}{f P^0}\delta(r-R(t))\right ]_{r} \nonumber\\
&-\frac{8\pi}{P^0}\frac{r^2 f\Pi_1}{(\lambda+1)\Lambda}\delta(r-R(t)) +\frac{1}{4 r (\lambda+1)\Lambda}\Big \{6(\lambda+1)\tilde{q}_{1R} \nonumber\\
&\qquad\qquad+[12M-r(4+\Lambda)]\tilde{q}_{2R} +2(r-6M)\tilde{q}_{1R,r}\Big \} \nonumber \\
&+\frac{1}{4  (\lambda+1)\Lambda} \left [ (\Lambda-2\lambda)\tilde{q}_{1R,r}+2r f(\tilde{q}_{2R,r}-\tilde{q}_{1R,rr})\right ] 
\\N'_4&=-4(\lambda+1)\frac{\tilde{h}^*_{0S}}{r^2}-4\left( \frac{M}{r}-1\right)\frac{\tilde{H}_{1S}}{r}+2f\frac{\partial \tilde{H}_{1S}}{\partial r}  -\frac{\partial I_S}{\partial t}  \\
N'_5&= \frac{M}{r^2}\tilde{H}_{0S}-2(\lambda+1)\frac{f \tilde{h}^*_{1S}}{r^2}+\left( \frac{2}{r}-\frac{3M}{r^2}\right)\tilde{H}_{2S}-\frac{2f}{r}\tilde{K}_S \nonumber\\
&+\frac{2(M-r)(1+\lambda)\tilde{q}_{1R}+r f(r(2+\Lambda)-2M)(\tilde{q}_{2R}-\tilde{q}_{1R,r})}{r^3(1+\lambda)\Lambda}  \nonumber\\
&+\frac{1}{2}f\frac{\partial (\tilde{H}_{0S}+\tilde{H}_{2S}-\tilde{K}_S)}{\partial r} -\frac{\partial \tilde{H}_{1S}}{\partial t} 
\end{align}
\begin{align}
N'_6&=\frac{2f}{r} \left [ \left ( \frac{M}{r}-1\right)\tilde{h}^*_{1S}+\left( \frac{\lambda}{2}\tilde{G}_{S}+\frac{\tilde{H}_{2S}-\tilde{H}_{0S}}{4}\right)r\right] \nonumber\\
&+f^2\frac{\partial \tilde{h}^*_{1S}}{\partial r}-\frac{\partial \tilde{h}^*_{0S}}{\partial t} \nonumber \\
&+f \Bigg \{ \frac{\tilde{q}_{2R}-\tilde{q}_{1R,r}}{2(\lambda+1)} \nonumber\\
&\quad\;\; +\left(1+r\frac{\partial}{\partial r}\right)\left[\frac{\tilde{q}_{1R}}{r\Lambda}-\frac{f}{(\lambda+1)\Lambda}\left(\tilde{q}_{2R}-\frac{\partial \tilde{q}_{1R}}{\partial r}\right)\right] \Bigg\}.
\end{align} 
\end{subequations}

We can pick the initial condition that $\tilde{G}_R=\tilde{h}^*_{1R}=\tilde{H}_{1R}=\tilde{h}^*_{0R}=I_R=\pi_2=0$. Eq.~(\ref{eqhareevo}) determines their evolution later on. Once $\tilde{G}_R$ and $\tilde{h}^*_{1R}$ are known, $\tilde{K}_R$ and $\tilde{H}_{2R}$ can be obtained using Eq.~(\ref{eqape}). $\tilde{H}_{1R}$ is just $I_R-\tilde{H}_{2R}-2\tilde{K}_R$ and then all effective fields for even parity are obtained following the above procedure. In reality, one may let the test particle freely evolve for a few cycles before turning on the  radiation reaction in order for the initial junk radiation to go away. Another subtlety here is although $\tilde{G}_R, \tilde{h}^*_{1R}, \tilde{I}_R,\tilde{H}_{1R},\tilde{h}^*_{0R}$ are all regular functions, $\pi_2$ may actually be divergent at $r=R(t)$. However, as long as the particle trajectory does not hit the grid point (which is the generic case and can be guaranteed by using some numerical algorithm), Eq.~(\ref{eqhareevo}) can still be used for harmonic gauge evolution.

\subsection{Monopole and dipole perturbations}

Although there is no wave equation in the monopole and dipole cases, regularization does involve these orders.  We need to carry out steps described in the above, simply ignoring the step of solving the wave equation. This is discussed by Detweiler and Poisson \cite{poisson2}, but unfortunately there is no known unique way to remove the singular piece of contribution from the $l=0,1$ modes. We will leave this for future investigation.

\mycomments{
\section{Monopole and dipole perturbations}
Analysis in previous section only dealt with $l\ge 2$ perturbations, and it has been well known that $l=0,1$ perturbations of Schwarzschild are not dynamical: at linear order, they do not have their own Hamiltonians, and can be determined solely from constraint equations and gauge-fixing functions.  More specifically, in vacuum, the $l=0$ perturbation is a static change of the mass of the black hole; the $l=1$ even party perturbation is a static change of the mass dipole moment, and hence its ``center-of-mass'' location in our coordinate system; the $l=1$ odd parity perturbation is a static change of the angular momentum of the black hole.  Here we explicitly deal with each of these modes, both in absence and presence of the point particle.  
\subsection{Monopole $(l=0)$ mode}
For $l=0$, there is only even-parity perturbations.  For perturbations of the 3-metric, we only have $H_2$ and $K$, while $G$ and $h_1^*$ both vanish due to the non-existence of the vector and tensor harmonics $Z$ and $V$.   In addition, one gauge transformation exists for this mode, and hence we only have one independent metric-perturbation function, which can then be determined by the Hamiltonian constraint.  This means we do not have to use the Hamiltonian at all to evolve metric perturbation. 
Now when we add the point particle into the picture, the Hamiltonian constraint will have a delta-function at the position of the particle.  This means for $r>R(t)$ and $r<R(t)$ we must have different solutions to the constraint equation, each with a gauge of our choice. Now, both these choices will enter the equation of motion of the particle and affect its coordinate evolution.  The questions are: (i) how should we appropriately choose the gauges?  and (ii) at the end, do we really have a $l=0$ self foce?
Apparently, at second order, the $l=0$ mode will have to serve as the intermediary that modifies the mass of the central black hole as it absorbed the GW --- but what would be the leading order?  I speculate the only thing is that it actually makes the out-side observer feel the total mass of 
\begin{equation}
\sim M + m - \frac{M m }{2R}
\end{equation}
instead of either $M$ or $M+m$. 
\subsection{Even-Parity Dipole   $(l=1)$ Mode}
In presence of particle, the joint effect of $l=0,1$, even-parity perturbations,  presumably ``resums'' into the Hamiltonian of the particle in such a way that it should be like an object with 
\begin{equation}
\mu = \frac{{m}{M}}{M+m}
\end{equation}
moving in the space-time of an object with mass $M+m$. 
\subsection{Odd-Parity Dipole  $(l=1)$ Mode}
}

\section{Conclusions and Discussions}\label{sec5}

 In this article we have taken a $3+1$ Hamiltonian approach toward the motion of a point particle around a Schwarzschild black hole. 
For the metric perturbation fields, we have simply adopted Moncrief's perturbative Hamiltonian (quadratic in these fields), and his canonical transformation to a new set of canonical coordinates and momenta which are either the Hamiltonian and momentum constraints themselves, their canonical conjugates, or gauge-invariant (see Sec.~\ref{subsec:dof}).   For the point particle, we have inserted its own Hamiltonian plus an interaction Hamiltonian, with the former describing geodesic motion and the latter describing both (i) the particle souring metric perturbations and (ii) the  metric perturbations acting back onto the particle.    We have obtained these equations of motion explicitly --- decomposed into $(l,m)$ components ($l\ge 2$) and even and odd parities.   For (i), the equations we obtain agree with the previous literature, obtaining wave equations for gauge-invariant functions that are sourced by the particle.  In this way, we have obtained self-consistent evolution equations for both the particle and the metric-perturbation fields.  In principle, depending on the lapse and shift functions we choose, these self-consistent equations can be written for any gauge.   The field equations we have will be in 1+1 ($t$ and $r$) dimensions, the gauge-invariant metric-perturbation fields are also decoupled from the rest of the fields.


As can be anticipated, these set of equations are singular due to the use of a point particle.  While we have not been able to find a stand-alone 3+1 approach for regularization, we have shown that existing regularization schemes can be adopted to our scheme.  The most straightforward approach would be to use the Detweiler-Whiting's singular-regular decomposition \cite{detweiler2}, combined with the Vega-Detweiler effective-source approach \cite{Ian1}.   In the case when we have a high order approximation of the singular field, one can (i) solve the wave equation for the even- and odd-parity effective gauge-invariant fields, and (ii) fix an algebraically simple gauge for the effective metric, obtain all effective metric components, and calculate generalized forces acting on the particle.    In this way, we will only have evolve one wave equation for each parity and each $l$ (and all  $m$'s can be taken care of simultaneously)  with an effective source, and use these waves, plus regularized gauge-fixing terms, to drive the motion of the particle in the self-consistent way.  However, since we have not carried out this computation explicitly, it is not   up to which level of approximation we shall require for the singular field --- although it may be substantially higher than what has been required before due to the multiple spatial derivatives used in defining the gauge-invariant quantities. 
 
In case the requirement for the singular field in (ii) turns out to be too high, we have proposed to replace (ii) by (ii'): choosing the Lorenz gauge for the full field, in which case it was known that the currently available approximations for the singular field is sufficient.  In this case, in addition to the two wave equations, we require 2 odd-parity gauge-fixing equations, and 6 even-parity gauge fixing equations.  This will be equivalent to decomposing  Vega-Detweiler's and Warburton et al.~\cite{warburton}'s 3+1 calculations into a 1+1 form.

For $l=0,1$ metric perturbation fields do not have gauge-invariant components: a canonical transformation exists to transform them into either the constraints or their canonical conjugates.   In this way, we only need step (ii) or (ii') in the above discussion --- which we have not explicitly carried out.

\begin{acknowledgements}
We thank Leor Barack, Tanja Hinderer, Chad Galley and Anil Zenginoglu for very helpful discussions. We also would like to thank Eric Poisson for providing Karl Martel's thesis and  Steven Detweiler for answering questions about effective source approach in self-force calculations. . This work is supported by NSF grants PHY-0555406, PHY-0956189, PHY-1068881, as well as the David and Barbara Groce startup fund at Caltech.
\end{acknowledgements}

\appendix
\section{various tensor spherical harmonics}\label{ap1}

Here we list the components of the vector and tensor harmonics $X^{lm}_A, Z^{lm}_A, W^{lm}_{AB},U^{lm}_{AB}, V^{lm}_{AB}$ in terms of scalar spherical harmonics and their partial derivatives. For the odd-parity vector harmonics $X^{lm}_A$, we have
\begin{align}
X^{lm}_{\theta} =-\frac{1}{\sin{\theta}}\frac{\partial Y^{lm}}{\partial \phi}, \, X^{lm}_{\phi} =\sin{\theta}\frac{\partial Y^{lm}}{\partial \theta}
\end{align}
For the even-parity vector harmonic $Z^{lm}_A$:
\begin{align}
Z^{lm}_{\theta} =\frac{\partial Y^{lm}}{\partial \theta}, \, Z^{lm}_{\phi} =\frac{\partial Y^{lm}}{\partial \phi}
\end{align}
For the identity tensor $U^{lm}_{AB}$:
\begin{align}
U^{lm}_{\theta\theta} = Y^{lm}, \quad  U^{lm}_{\phi\phi} =\sin^2{\theta} Y^{lm}, \quad  U^{lm}_{\theta\phi}=U^{lm}_{\phi\theta}=0\,.
\end{align}
For the even-parity, symmetric trace-free (STF) tensor harmonic $V^{lm}_{AB}$:
\begin{subequations}
\begin{align}
V^{lm}_{\theta\theta} &= \left [ \frac{\partial^2}{\partial \theta^2}+\frac{1}{2}l(l+1)\right ]Y^{lm}  \\
 V^{lm}_{\phi\phi} &= \left [ \frac{\partial^2}{\partial \phi^2}+\cos{\theta}\sin{\theta}\frac{\partial}{\partial \theta}+\frac{1}{2}l(l+1)\sin^2{\theta}\right ]Y^{lm}  \\
V^{lm}_{\theta\phi} &= V^{lm}_{\phi\theta} = \left [ \frac{\partial^2}{\partial \theta \partial \phi} -\frac{\cos{\theta}}{\sin{\theta}} \frac{\partial}{\partial \phi}\right ]Y^{lm}\,.
\end{align}
\end{subequations}
And finally for the odd-parity tensor harmonic $W^{lm}_{AB}$:
\begin{align}
W^{lm}_{\theta\theta} &= -\frac{1}{\sin{\theta}}\left [ \frac{\partial^2}{\partial \theta\partial \phi} -\frac{\cos{\theta}}{\sin{\theta}} \frac{\partial}{ \partial \phi}\right ]Y^{lm} \nonumber \\
W^{lm}_{\phi\phi} &=\left [\sin{\theta} \frac{\partial^2}{\partial \theta\partial \phi} -\cos{\theta} \frac{\partial}{ \partial \phi}\right ]Y^{lm} \nonumber \\
W^{lm}_{\theta\phi} &= W^{lm}_{\phi\theta} = \frac{1}{2}\left [ \sin{\theta}\frac{\partial^2}{\partial \theta^2 } -\frac{1}{\sin{\theta}}\frac{\partial^2}{\partial \phi^2 }-\cos{\theta}\frac{\partial}{\partial \theta}\right ]Y^{lm}
\end{align}

\section{Conversion of Fields for Even Parity Perturbations}\label{ap2}
The even parity perturbation quantity $K, H_2, h^*_1, G$ and $q_1, q_2, q_3, q_4$ are related to each other by
 \begin{eqnarray}\label{eqape}
 K &=& \frac{q_1}{r \Lambda}-\frac{f}{(\lambda+1)\Lambda}(q_2-q_{1,r}) - r f \left ( q_{3,r}-\frac{2 q_4}{r^2}\right )\nonumber \\
 H_2   &=& \frac{(\Lambda-1)(\Lambda-2-2\lambda)}{2 f r \Lambda}q_1+\frac{q_2-q_{1,r}}{2(\lambda+1)} \nonumber \\
 &+& r \left [ \frac{q_1}{r \Lambda}-\frac{f}{(\lambda+1)\Lambda}(q_2-q_{1,r})\right ]_{,r}\nonumber \\
 &-& f[r^2 q_{3,r}-2q_4]_{,r}-\frac{M}{r^2}(r^2q_{3,r}-2q_4) \nonumber \\
 &=&\frac{q_2-q_{1,r}}{2(\lambda+1)}+ (r K)_{,r}-K-\frac{1}{r}(1-\frac{3M}{r})\left[ r^2 q_{3,r}-2 q_4\right ] \nonumber \\
  G &=& q_3 \nonumber \\
 h^*_1 &=& q_4 
 \end{eqnarray}

\end{document}